\documentclass[12pt,preprint]{aastex}
\usepackage{graphicx, apjfonts}
\usepackage{epsf}
\usepackage{natbib}

\shortauthors{Robaina et al.}
\shorttitle{Galaxy Interactions and Star Formation}

\newcommand{\peg}{{\sc P\'egase }}

\newcommand{\um}{$\mu$m }

\newcommand{\avg}[1]{\left< #1 \right>} 

\begin{document}


\title{Less than 10 percent of star formation in $z \sim 0.6$ massive galaxies is triggered by major interactions}

\author{Aday R. Robaina$^1$,
Eric F.\ Bell$^1$,
Rosalind E. Skelton$^1$,
Daniel H. McIntosh$^{2,3}$,
Rachel S. Somerville$^4$,
Xianzhong Zheng$^5$,
Hans-Walter Rix$^1$,
David Bacon$^6$, 
Michael Balogh$^7$,
Fabio D. Barazza$^8$,
 Marco Barden$^9$, 
Asmus B\"ohm$^{10}$,
John A.R. Caldwell$^{11}$,
Anna Gallazzi$^1$, 
Meghan E.\ Gray$^{12}$, 
Boris H\"aussler$^{12}$,
Catherine Heymans$^{13}$, 
Knud Jahnke$^1$, 
Shardha Jogee$^{14}$, 
Eelco van Kampen$^{9,20}$,
Kyle Lane$^{12}$,
Klaus Meisenheimer$^1$,
Casey Papovich$^{15}$,
Chien Y. Peng$^{16}$,
Sebasti\'an F. S\'anchez$^{17}$, 
Ramin Skibba$^1$,
Andy Taylor$^{18}$,
Lutz Wisotzki$^{10}$,
Christian Wolf$^{19}$}

\affil{$^1$ Max-Planck-Institut f\"ur Astronomie, K\"onigstuhl 17, D-69117 Heidelberg, Germany; \texttt{arobaina@mpia.de}\\
$^2$ Department of Astronomy, University of Massachusetts, 710
  North Pleasant Street, Amherst, MA 01003, USA\\
$^3$ Department of Physics, University of Missouri-Kansas City,
  MO 64110, Kansas City,  USA \\
$^4$ Space Telescope Science Institute, 3700 San Martin Dr., Baltimore, MD 21218, USA\\
$^5$ Purple Mountain Observatory, Chinese Academy of Sciences, Nanjing 210008, PR China\\ 
$^6$ Institute of Cosmology and Gravitation, University of Portsmouth, Hampshire Terrace, Portsmouth PO1 2EG\\
$^7$ Department of Physics and Astronomy, University Of Waterloo,
  Waterloo, Ontario, Canada N2L 3G1\\
$^8$ Laboratoire d'Astrophysique, \'Ecole Polytechnique
  F\'ed\'erale de Lausanne (EPFL), Observatoire, CH-1290 Sauverny, Switzerland\\
$^9$ Institute for Astro- and Particle Physics, University of Innsbruck, Technikerstr. 25/8, A-6020 Innsbruck, Austria\\
$^{10}$ Astrophysikalisches Institut Potsdam, An der Sternwarte 16,
  D-14482 Potsdam, Germany\\
$^{11}$ University of Texas, McDonald Observatory, Fort Davis, TX
  79734, USA\\
$^{12}$ School of Physics and Astronomy, University of Nottingham,
     Nottingham NG7 2RD, UK\\
$^{13}$ Department of Physics and Astronomy, University of British Columbia, 6224 Agricultural Road, Vancouver, Canada V6T 1Z1\\
$^{14}$ Department of Astronomy, University of Texas at Austin, 1 University Station, C1400 Austin, TX 78712-0259, USA\\
$^{15}$ Steward Observatory, The University of Arizona, 933 North
Cherry Avenue, Tucson, AZ 85721, USA\\
$^{16}$ NRC Herzberg Institute of Astrophysics, 5071 West Saanich
  Road, Victoria, Canada V9E 2E7\\
$^{17}$ Centro Hispano Aleman de Calar Alto, C/Jesus Durban Remon
  2-2, E-04004 Almeria, Spain\\
$^{18}$ The Scottish Universities Physics Alliance (SUPA), Institute for Astronomy, University of Edinburgh, Blackford Hill,
    Edinburgh, EH9 3HJ, UK\\
$^{19}$ Department of Physics, Denys Wilkinson Bldg., University
of Oxford, Keble Road, Oxford, OX1 3RH, UK\\
$^{20}$ European Southern Observatory, Karl-Schwarzschild-Str. 2, D-85748 Garching,  
Germany
}

\begin{abstract}
  Both observations and simulations show that major tidal interactions or mergers 
between
gas--rich galaxies can lead to intense bursts of star
formation.  Yet, the {\it average} enhancement in star formation rate (SFR)
in major mergers and the contribution of such events to the cosmic SFR are
not well estimated. Here we use photometric redshifts, stellar masses and
  UV SFRs from COMBO-17, 24\um SFRs from {\it Spitzer} and morphologies from
  two deep Hubble Space Telescope (HST) cosmological survey fields (ECDFS/GEMS
  and A901/STAGES) to study the enhancement in SFR as a function of projected
  galaxy separation. We
apply two-point projected correlation function techniques, which we augment with morphologically-selected
very close pairs (separation $<2$\arcsec) and merger remnants from the HST imaging. Our
analysis confirms that the most intensely star-forming systems are indeed interacting or merging.
Yet, for massive ($M_* \ge 10^{10}M_\sun$) star-forming galaxies
at $0.4<z<0.8$, we find  that the SFRs of galaxies undergoing a major
interaction (mass ratios $\leq1:4$ and separations $\leq 40$\, kpc) are only
$1.80\pm 0.30$ times higher than the SFRs of non-interacting galaxies when
averaged over all interactions and all stages of the interaction, in good 
  agreement with other observational works.
 Our results also agree with hydrodynamical simulations
of galaxy interactions, which produce some 
mergers with large bursts of star formation on $\sim 100$\,Myr 
timescales, but only a modest SFR enhancement when averaged over the entire merger
timescale. We demonstrate that these results imply that only $\lesssim 10\%$ of 
star formation at $0.4\le z \le 0.8$ is {\it triggered directly} 
by major mergers and interactions; these events are {\it not} important
factors in the build-up of stellar mass since $z=1$.

\end{abstract}

\keywords{galaxies: general ---  
galaxies: evolution --- galaxies: star formation --- galaxies: interactions ---
infrared: galaxies }

\section{Introduction}
Observational evidence from a variety of angles indicates that galaxy
interactions and mergers of galaxies can lead to dramatically-enhanced
star formation \citep{sanders88, barton2000, lambas, barton07}.  This
appears to hold true at all redshifts where one can recognize mergers
through galaxy morphologies ($z \la 1$ with rest-frame optical
morphologies; \citealp{melbourne, hammer05}; $1 \la z \la 3$ using less
certain UV morphologies \citealp{chapman}). Ultra-luminous infrared galaxies (ULIRGs), representing the
highest-intensity star formation events at low redshifts, are almost
invariably hosted by merging galaxies \citep{sanders88}. For a number
of applications the quantity of interest is the {\it average}
enhancement in star formation (SF) triggered by merging (ensemble average over the
population of major mergers/interactions, or equivalently, temporal
average over major merger events during a merger lifetime),
not the high-intensity tail \citep[e.g.,][]{barton2000, lambas, lin,
  li, shardha}. \cite{barton07} carefully quantified the star formation rate (SFR)
enhancement in mergers in low-mass halos at low redshift, using the
Two-Degree Field Galaxy Redshift Survey \citep{colless}. They found
that roughly 1/4 of galaxies in close pairs (separated by $<50$\,kpc) in
low-mass halos with $M_{b_J} < -19$ have SFR enhancements of a factor
of five or more\footnote{This corresponds roughly to a mass cut of $5
  \times 10^9 M_{\sun}$, assuming a stellar $M/L_{b_J} \sim 1$,
  appropriate for a star-forming blue galaxy with a \cite{chabrier03}
  stellar IMF.}.

It has also been noted that the strong decrease of the cosmic SFR
density between $z=1$ and $z=0$ \citep[e.g., ][]{lilly, madau,
  hopkins04, lefloch} was not dissimilar from the relatively rapid
drop in merger rate inferred (at that time) from close pairs and
morphologically-selected mergers \citep{lefevre}.  If much of the star
formation at $z>0.5$ were triggered by merging, the apparent
similarity in evolution between SFR and merger rate would be a natural
consequence.  More recently, studies of the fraction of star formation
in morphologically-selected interacting and merging galaxies at
intermediate redshifts $z<1$ have demonstrated that, in fact, the bulk
of star formation is in quiesciently star-forming disk-dominated
galaxies \citep{hammer05, wolf05, bell05, shardha}.

Similarly, it has long been argued that early-type (elliptical and
lenticular) galaxies are a natural outcome of galaxy mergers
\citep[e.g., ][]{toomre,ss}. In any hierarchical cosmogony mergers are
expected to play a large role; a wide range of work --- observations
of the increasing number density of non-star-forming early-type
galaxies from $z=1$ to the present \citep{bell04, brown07, faber07},
the kinematic and stellar populations of local early-type galaxies
\citep{trager00, emsellem04}, or the joint evolution of the stellar
mass function and star formation rates of galaxies \citep{bell07,
  walcher08, perez08} --- has given support to the notion that at
least some of the early-type galaxies assembled at $z<1$ have done so
through galaxy merging. In such a picture, the average SFR enhancement
from merging is of interest for interpreting the SF and chemical
enrichment history of early-type galaxies, inasmuch as it gives an
idea of what kind of fraction of stars in present-day early-type
galaxies we can expect to have formed in the burst mode, and what
fraction we can expect to have formed in a quiescent mode in the
progenitor galaxies.

Direct observational constraints on the enhancement in SFR caused by
merging provide an important calibration for modeling triggered star
formation in cosmologically-motivated galaxy formation
models. Hydrodynamic simulations of interacting galaxies in which gas
and star formation are explicitly modeled have demonstrated that
torques resulting from the merger can efficiently strip gas of its
angular momentum, driving it to high densities and leading to
significant enhancement in star formation
\citep[e.g.,][]{BH96,MH96,cox06,cox08,dimatteo07}. However,
state-of-the-art cosmological simulations lack the dynamic range to
accurately simulate the internal structure of galaxies in significant
volumes, so estimates of the global implications of merger-driven star
formation enhancement have had to rely on semi-analytic calculations
\citep[e.g.][]{spf01,baugh05,somer08}. Furthermore, as the progenitor properties play a key role in the
simulated SFR enhancements \citep[e.g.,][]{dimatteo07, cox08},
inaccurate progenitor property values (e.g., incorrect gas fraction or
internal structure) will lead to incorrect estimates for the average
fraction of SF in mergers even if the SF in each individual merger
were modeled perfectly.

Therefore, to constrain galaxy evolution models and to understand the
physical processes responsible for the main mode of star formation at $z<1$, it is of interest to determine observationally the
typical enhancement\footnote{When we refer to SFR enhancement, we
  define this as the ratio of SFR in some subsample (e.g., close
  pairs) to the average SFR of all systems in that mass bin.} in SFR
averaged over the duration of the entire major (stellar-mass ratio between 1:1 and 1:4) galaxy merger or
interaction and to constrain
the overall fraction of SF triggered by mergers/interactions at
intermediate redshift. In a companion paper \citep{shardha} we focus on the rate of merging
and also present a preliminary exploration of the average change in the SFR caused
by late-stage major and minor merging, finding an average mild
enhancement within the restrictions imposed by the sample size. In
this paper we present a statistically-robust analysis of the
properties of star-forming galaxies at $0.4<z<0.8$ including all
relevant merger phases and aimed at providing a satisfactory answer to
two key questions.  What is the average enhancement in star formation
rate as a function of galaxy pair separation compared to their SFR
before the interaction?  What fraction of star formation is directly
triggered by major mergers and interactions?

There are a number of conceptual and practical challenges in such an
experiment.  Enhancements in SFR produce both a boost in luminosity,
but also increase dust content and extinction. At a minimum, one
therefore needs dust-insensitive SFR indicators. In addition,
simulations have indicated that SF can be enhanced at almost all
phases of an interaction from first passage through to after
coalescence \citep[e.g.,][]{BH96, dimatteo07}; although close pairs will
  inevitably include some fraction of galaxies before first pass and galaxies
  with unbound orbits. Therefore, an analysis
needs to include both close pairs of galaxies (those before coalescence) and morphologically-classified mergers
(primarily those near or after coalescence).  Morphological
classification is not a straightforward art \citep[see][for a
  comparison between automated classifications and visual
  morphologies]{shardha}, even in ideal cases \citep{lisker}.
Finally, galaxy mergers are rare and short-lived, necessitating large
surveys to yield substantive samples of mergers.

In this work, we address these challenges as far as possible (see also
\citealp{lin} and \citealp{li}).  We use estimates of redshift and
stellar mass from the COMBO-17 survey \citep{wolf03, borch} to define
and characterize the sample. Stellar mass selection should limit the
effect of enhanced star formation and dust content on the sample
definition.  We use SFR indicators that are constructed to be dust
extinction insensitive, by combining ultraviolet (UV; direct,
unobscured light from young stars) and infrared (IR; thermal emission
from heated dust, powered primarily by absorption of UV light from
young stars) radiation \citep{bell05}.  Finally, we study a very
well-characterized sample of galaxy pairs at $0.4<z<0.8$ using
weighted projected two-point correlation functions \citep{ramin,li},
supplementing them at very small separations $\la 15$\,kpc with very
close pairs or merger remnants morphologically selected from two wide
HST mosaics, GEMS \citep{rix04} and STAGES \citep{gray}, in an attempt
to account for all stages of galaxy interactions.

The plan of this paper is as follows.  In \S \ref{data} we discuss the
data and the methods used to estimate the stellar masses and the
SFRs. In \S \ref{sample} we describe the sample selection and the
method used for the analysis. In \S \ref{results} we present our
estimates of the enhancement in SFR as a function of projected
separation.  In \S \ref{disc}, we compare with previous observations,
constrain the fraction of SF triggered by major mergers and
interactions at $0.4<z<0.8$, and compare with simulations of galaxy
merging.  Finally in \S \ref{conc} we summarize the main findings of
this paper.  All the projected distances between the pairs used here
are proper distances.  We assume $H_{0}=70$\,km\,s$^{-1}$\,Mpc$^{-1}$,
$\Omega_{\Lambda 0}=0.7$ and $\Omega_{\rm m0}=0.3$.

\section{The Data} \label{data}

\subsection{COMBO-17. Redshifts and stellar masses}

COMBO-17 has to date fully surveyed and analyzed three fields 
to deep limits in 5 broad
and 12 medium pass-bands (Extended Chandra Deep Field South (ECDFS), A901/2 and S11, see \citet{wolf03} and \citet{borch}). Using galaxy, star and
quasar template spectra, objects are classified and redshifts assigned for
$\sim$ 99\% of the objects to a limit of $m_{R} \sim$ 23.5 \citep{wolf04}. 
The photometric redshift errors can be described as
\begin{equation}
\frac{\sigma_z}{1+z} \sim 0.007 \times [1 + 10^{0.8(m_R-21.6)}]^{1/2}, \label{eqn:dz}
\end{equation}
and rest frame colors and absolute magnitudes are accurate to $\sim$ 0.1 mag 
(accounting for distance and $k$-correction uncertainties). The astrometry
is accurate to $\sim 0.1$\arcsec and the average seeing is
  0.7$\arcsec$. It is worth noting that Eq.~\ref{eqn:dz} leads to typical
  redshift errors of $\sigma_z\simeq 0.01$ for bright ($m_R < 21$) and
  $\sigma_z\simeq 0.04$ for faint ($21< m_R < 23.5$) galaxies in the
  $0.4<z<0.8$ interval.

 The stellar masses were estimated in COMBO-17 by \cite{borch} using the
 17-passband photometry in conjunction with a non-evolving template library
 derived using the \peg stellar population model \citep[see][]{fioc97,fioc99} and a
 \cite{kroupa93} initial mass function (IMF). Note that the results assuming a
 \cite{kroupa01} or a \cite{chabrier03} 
 IMF yield similar stellar masses to within $\sim$
 10\%. The reddest templates have smoothly-varying exponentially-declining star
 formation episodes, intermediate templates have a contribution from 
 a low-level constant level of star formation, while the bluer templates have a recent burst of
 star formation superimposed. 

 The masses are consistent with those using M/L estimates based on a single
 color \citep[e.g., ][]{bell03}. Random stellar mass errors are $< 0.3$ dex on a
 galaxy-by-galaxy basis, and systematic errors in the stellar masses were
 argued to be at the 0.1 dex level \citep[see][for more details]{borch}.
 \cite{bell01} argued that galaxies with large bursts of recent 
star formation could produce stellar M/L values  
at a given color that are lower by up to 0.5 dex; this uncertainty is more
 relevant in this work than is often the case.  While this will inevitably remain an uncertainty here, 
we note that the \cite{borch} templates do include bursts explicitly, 
thus compensating for the worst of the uncertainties introduced by bursting 
star formation histories. In \S\ref{uncertainties} we will explicitly study the
  impact that such uncertainties have on our results.

 In what follows, we use COMBO-17 data for two fields: the ECDFS and Abell 901/902 fields, because of their
complementary data: deep HST/ACS imaging from the GEMS and STAGES 
projects respectively (allowing an investigation of morphologically-selected
merger remnants and very close pairs), and deep 24{\micron} imaging
from the MIPS instruments on board {\it Spitzer}, required to measure obscured SF.  

\subsection{GEMS and STAGES HST imaging data}

 F606W (V-band) imaging from the GEMS and STAGES surveys provides
 0.1\arcsec \,resolution images for our sample of COMBO-17 galaxies. Using the Advanced Camera
 for Surveys \citep[ACS;][]{ford} on board the Hubble Space Telescope (HST),
 areas of $\sim 30\arcmin \times 30\arcmin$ in each of 
 the ECDFS and the A901/902
 field have been surveyed to a depth allowing galaxy detection to a
 limiting magnitude of $m_{lim}^{AB}(F606W)=28.5$ \citep{rix04,gray,caldwell}. 
 These imaging data are later used to visually classify galaxies, allowing
 very close pairs (separations $<2"$) and merger remnants to be included
in this analysis. We choose not to use F850LP HST data available for the GEMS
 survey in order to be consistent in our classification between the two fields
 (only F606W is available from STAGES).

\subsection{MIPS 24 \um, total infrared emission and star formation rates}\label{mips}

The IR observatory Spitzer has surveyed two of the COMBO-17 fields: a $1\arcdeg
\times 0.\arcdeg5$ scan of the ECDFS (MIPS GTO), and a 
similarly-sized field around
the Abell 901/902 galaxy cluster (MIPS GO-3294: PI Bell). The final images have
a pixel scale of $1\arcsec.25$/pixel and an
image PSF FWHM of $\simeq 6\arcsec$. Source detection and photometry are
described in depth in \cite{papovich04} and catalogue matching in
\cite{bell07}\footnote{In this paper, we are interested in SFR enhancements in
close pairs of galaxies, where the closest pairs may fall within a single {\it
Spitzer}/MIPS PSF. Accordingly, in this work we choose to explore the total
SFR in the pair (avoids deblending uncertainties) rather than the
individual SFR occuring in both galaxies.}. Based on those works, we estimate that our source detection is
80\% complete at the $5 \sigma$ limit of 
$83\mu$Jy in the 24\um data in the ECDFS for a
total exposure of $\sim$ 1400 s pix$^{-1}$. The A901/902 field 
has similar exposure time, but owing to higher (primarily zodiacal) background
the $5 \sigma$ limit (80\% completeness) is 97$\mu$Jy, with lower 
completeness of 50\% at 83$\mu$Jy.  We use both catalogs to a limit
of 83$\mu$Jy. 

To include both obscured and unobscured star formation into the estimate of
the SFR of galaxies in our sample, we combine UV emission 
with an estimate of the total IR 
luminosity in concert. As 
the total thermal IR flux in the 8--1000\um range is 
observationally inaccessible for almost all galaxies in our 
sample, we have instead estimated total IR luminosity 
from the observed 24{\micron} flux, corresponding to 
rest-frame 13--17\um emission at the redshifts of interest $z=0.4-0.8$. 
For this exercise, we adopt a Sbc template from the \cite{dev99}
SED library \citep{zheng07a, bell07}.  
The resulting IR luminosity is accurate to a factor of $\lesssim 2$: 
local galaxies with IR luminosities in excess of $>10^{10} L_{\sun}$ 
show a tight correlation between rest-frame
12--15\um luminosity and total IR luminosity
\citep{spi95,cha01,rou01,papovich02} with a scatter of $\sim$ 0.15 dex.
Furthermore, \cite{zheng07a} have stacked luminous 
($L_{TIR}\ga 10^{11}{\rm L_{\sun}}$) $z \sim 0.7$ galaxies
at 70\um and 160\um, finding that their average spectrum 
is in good agreement with the Sbc template from 
\cite{dev99}, validating at least on average our choice
of IR SED used for extrapolation of the total IR luminosity.

We estimate the SFR by using both directly observed UV-light from massive
stars and dust-obscured UV-light measured from the mid-infrared. As in
\cite{bell05} we estimate the SFR $\psi$ by means of a calibration derived
from \peg synthetic models assuming a 100 Myr-old stellar population with
constant SFR and a \cite{chabrier03} IMF: 

\begin{equation}
\psi / ({\rm M_{\sun}\,yr^{-1}}) = 9.8 \times 10^{-11} \times
       (L_{\rm TIR} + 2.2L_{\rm UV}).  \label{eqn:sfr}
\end{equation}

Here $L_{\rm TIR}$ is the total IR luminosity and $L_{\rm UV} = 1.5 \nu l_{\nu,2800}$ is a rough
estimate of the total integrated 1216{\AA}--3000{\AA} UV luminosity. This
UV luminosity has been derived from the 2800{\AA} rest-frame luminosity from
COMBO-17 $l_{\nu,2800}$. The factor of 1.5 in the 2800{\AA}-to-total UV
conversion accounts for UV spectral shape of a 100 Myr-old population with
constant SFR, and the UV flux is multiplied by 2.2 to account for the light
emitted longwards of 3000{\AA} and shortwards of 1216{\AA} by the unobscured
stars belonging to the young population.

For all galaxies detected above the 83$\mu$Jy limit, we have used 
the IR and UV to estimate the total SFR.  For galaxies undetected
at 24{\micron}, or detected at less than 83$\mu$Jy, we use instead
UV-only SFR estimates.  

\subsubsection{IR emission from AGN-heated dust}

Possible contamination of mid-IR-derived SFRs from AGN heated dust is often
addressed by estimating the fraction of star formation held in X-ray detected sources.
In our case $< 15\%$ of the star forming galaxy sample were detected in X-rays, in 
good agreement with the results found by
i.e. \cite{silva04} or \cite{bell05}. 

Yet, there are two limitations of this estimate. Firstly, 
this does not account for any contribution from X-ray undetected
Compton-thick AGN, which could drive up the expected contribution 
from AGN in our sample.  For example, applying an $m_{R}=24$ cut to the
sample of \cite{alonso06}, we estimate the fraction of X-ray undetected AGN
to be $\sim 30\%$, while \cite{risaliti} find $\sim 50\%$ of local AGN to be
Compton thick. On this basis, it is conceivable that up to 30\%
of 24{\micron} luminosity is from galaxies with AGN\footnote{Although 
note that in a recent investigation of X-ray undetected 
IR-bright galaxies in the CDFS, \cite{lehmer} found
that radio-derived (1.4GHz) SFRs agree with the UV+IR-derived ones. This
implies that the relative strength of any AGN component 
is not dominant when compared to the host galaxy.}
.

Secondly, even in galaxies with AGN, not all of the IR emission will
come from the AGN. Although the data does not currently exists to answer
  this question conclusively, it is possible to make a rough estimate of the
effect. In order to estimate the fraction of mid-IR 
light that comes from the AGN (as opposed to star formation in the host), 
we have made use of the results of \cite{Cris}, who
attempted to structurally decompose mid-infrared imaging from 
\emph{Infrared Space Observatory} 
for a sample of both Seyfert 1 and 2 AGN in the local Universe, some of
which are very highly-obscured in X-rays. 
Analyzing the results in Tables 2 and 3 of \cite{Cris},
we have found that only a small fraction of the IR radiation 
at $\sim 10${\micron} (in this paper we work at rest-frame 13-17\um) comes from the central 
parts of the galaxies in the Seyfert 2
population, finding a total contribution of:
\begin{equation}
\frac{F_{IR}^{AGN}}{F_{IR}^{total}}=0.26 \pm 0.02.
\end{equation}
This result should be viewed as indicative only: obviously, 
the systems being studied will be different in detail from those
in our sample.  Furthermore, the 10{\micron} luminosities 
of the nuclei will be preferentially affected by silicate absorption, making 
it possible that our value of $F_{IR}^{AGN}/F_{IR}^{total}$ is a 
lower limit.  

Despite the various levels of uncertainty, taking the different 
lines of evidence together demonstrates that $\la 30$\% of the
IR luminosity in our sample comes from systems that may host an AGN, 
and that it is likely that $<10$\% of the IR luminosity of 
our sample is powered by accretion onto supermassive black holes.
Given the other uncertainties
in our analysis, we choose to neglect this source of error in what follows. 

\section{Sample selection and method}   \label{sample}

The goal of this paper is to explore the star formation rate
in major mergers between massive galaxies, from the pre-merger
interaction to after the coalescence of the nuclei. We chose a stellar mass--limited 
sample with $M_{\star}\ge10^{10}M_{\sun}$ in the redshift slice of $0.4< z\le 0.8$ 
(see Fig. 1). This roughly corresponds to $M_V=-18.7$ for galaxies in the
red sequence and $M_V=-20.1$ for blue objects. We only included
galaxies that fall into the footprint of both the ACS surveys GEMS and STAGES and
of existing 
{\it Spitzer} data. These criteria resulted in a final sample of 2551 galaxies.

Given the flux limit $m_R\lesssim 23.5$ for which COMBO-17 has reasonably
complete redshifts \citep{wolf04} we are complete for $M_*>10^{10}M_\sun$ blue 
cloud galaxies over the entire redshift range $0.4<z<0.8$. For 
red sequence galaxies, the sample becomes somewhat incomplete 
at $z>0.6$, and at $z=0.8$, the limit is closer to 
$2 \times 10^{10} M_{\sun}$. 
We chose to adopt a limit of $10^{10}M_{\sun}$ in what follows, 
despite some mild incompleteness in the red sequence, for two 
reasons.  First, adopting a cut of $2 \times 10^{10} M_{\sun}$ 
across the whole redshift range reduces the sample 
size by a factor of 30\%, leaving too small a sample for the proposed
experiment. Second, the vast
majority of the star forming galaxies are blue cloud galaxies ($83\%$ of the star formation
is occuring in blue galaxies), making the modest incompleteness 
in the red sequence of minor importance. 

\begin{figure}[h!]

\begin{center}
\includegraphics[width=15cm,height=9cm]{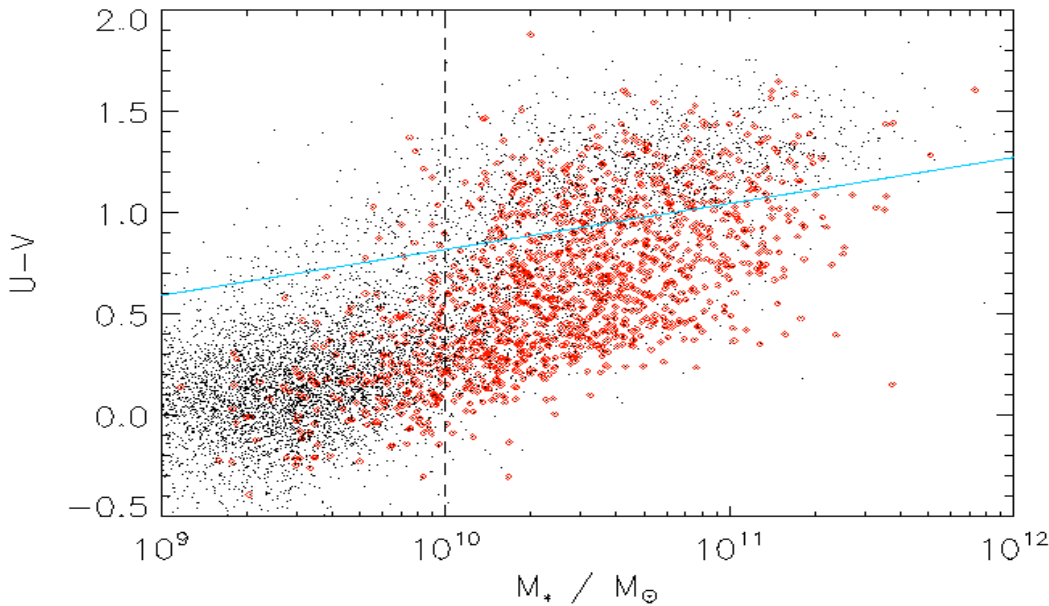}
\caption{\label{fig:cut} Stellar mass vs. color distribution of COMBO-17 selected galaxies in the ECDFS
and A901/2 field with $0.4<z<0.8$. 
The vertical line shows the mass
limit $M_{\star}>10^{10} M_{\sun}$ used to select our sample. This mass
selected sample is complete except for red sequence galaxies at $z>0.6$. The blue
line shows the cut used to separate red sequence and blue cloud
galaxies. Red symbols denote 24 \um detected galaxies with >$83 \mu$\,Jy.
}
\end{center}
\end{figure}

Later, we will use a subsample composed of star forming galaxies. We will
refer to 'star formers' as galaxies defined by having either blue optical 
colors or having been detected 
in the MIPS 24\um band. We select optically-blue galaxies
adopting a stellar mass-dependent 
cut in rest-frame $U-V$ color, following 
\cite{bell07}: $U-V\ga 1.06 -0.352z + 0.227(\log_{10} M_* -
10)$\footnote{Due to minor magnitude and color calibration differences between the two fields, the red
  sequence cut is slightly field dependent, with the intercept at
  $10^{10}M_{\sun}$ and z=0 being U-V=1.01 and 1.06 for the ECDFS and the
  A901/902 fields.} (see Fig.~\ref{fig:cut}).
We include all objects detected above the 24{\micron} limit 
of 83$\mu$\,Jy as star-forming.

In order to track star formation in very close pairs ($<2\arcsec$ and hence
unresolved by the ground-based COMBO-17 data) and merger remnants, we
include only merging systems (from the ACS data) with $M_*>2\times 10^{10}
M_{\sun}$: i.e. the minimum possible mass for a merger between two galaxies in
our sample. 

\subsection{Projected correlation function}

The correlation function formalism is a convenient and powerful tool to
characterize populations of galaxy pairs \citep[e.g.][]{davis, BK}.
Here, we use weighted projected two-point correlation functions because redshift 
uncertainties (1-3\%) from COMBO-17 translate to line-of-sight distance errors
 of $\sim100$\, Mpc, necessitating the use of projected correlation functions
to explore the properties of close physical pairs of galaxies
\citep{bell06}. For our sample at hand, we estimate the weighted (or marked) two-point
correlation function \citep{boerner, BK, ramin, ramin08}, using both the SFR
and the specific SFR (SFR per unit stellar mass) as the weight. 
 
The projected correlation function $w(r_P)$ is the integral along the line of sight
of the real-space correlation function:

\begin{equation}
w(r_{p})=\int^{\infty}_{-\infty} \xi([r_p^2+\pi^2]^{1/2})d\pi,
\end{equation}
where $r_{p}$ is the distance between the two galaxies projected on the plane
of sky and $\pi$ the line-of-sight separation. A simple estimator for this
unweighted correlation function is $w(r_{p})=\Delta(DD/RR-1)$, where $\Delta$
is the path length being integrated over, $DD(r_P)$ is the histogram of separations
between real galaxies and $RR(r_P)$ is the histogram of separations between galaxies
in a randomly-distributed catalogue (this is the same estimator used in
\citealp{bell06}). Basically, the aim is
  to find the excess probability (compared to a random distribution) of
  finding a galaxy at a given distance of another galaxy. This estimator
  accomplishes that by subtracting the random probability of finding two
  galaxies at a given separation from the probability in the real data sample
  and normalizing to the probability in the random case. Other estimators (i.e. $\Delta
[(DD-DR)/RR]$ or $\Delta[(DD-2DR+RR)/RR]$) for the 2-point correlation
function give results different by $<5\%$ (less than other sources of uncertainty). Thus:
\begin{displaymath}
DD(r_P)=\small{\sum_{ij}}D_{ij}
\end{displaymath}
\begin{displaymath}
RR(r_P)=\small{\sum_{ij}}R_{ij}
\end{displaymath}
where the sum is over all non-repeated pairs in the sample, and $D_{ij}$
  ($R_{ij}$) equals 1 only if the pair selection criteria are
  satisfied in the real
(random) galaxy catalogue, and is equal to 0 otherwise. The first criterion is that the
  stellar-mass ratio falls between 1:1 and 1:4. We further only allow a maximum redshift
difference $\Delta z=\Delta=\sqrt{2} \sigma_{z}$, where $\sigma_{z}$ is the error in
redshift of the primary galaxy (see Equation ~\ref{eqn:dz}), and, depending on
  the case, either the primary or both galaxies in the pair have to be star
  formers (see \S \ref{results}).

We can then study the possible enhancement of (specific) star formation rate
by means of a projected marked (or weighted) correlation function, which can be defined:
\begin{equation}
E(r_{p})=\frac{1+W(r_{p})/\Delta}{1+w(r_{p})/\Delta},
\end{equation}
where $W(r_{p})=\Delta(PP/PP_{R}-1)$ and,
\begin{displaymath}
PP(r_P)=\small{\sum_{ij}}P_{ij}D_{ij}
\end{displaymath}
\begin{displaymath}
PP_{R}(r_P)=\small{\sum_{ij}}P_{ij}R_{ij}.
\end{displaymath}
$P_{ij}$ is the mark (or weight). We adopt two different weights $P_{ij}$ in what follows, one is the
SFR of the pair of galaxies:
\begin{displaymath}
P_{ij} = S_{ij}=SFR_{ij}=SFR_{i}+SFR_{j},
\end{displaymath}
and the other is the specific SFR of the galaxy pair:
\begin{displaymath}
P_{ij} = s_{ij}={\rm Specific}\, SFR_{ij}=\frac{SFR_{i}+SFR_{j}}{M_{\star,i}+M_{\star,j}}.
\end{displaymath}
Then, the estimator that we use for $E(r_{p})$ is:
\begin{equation}
E(r_{p})=\frac{PP/DD}{\langle P_{ij}  \rangle},
\end{equation}
where $\langle P_{ij}\rangle$ is the average value of the weight used
  (SFR, or specific SFR) across the sample. It is worth noting that with our
  definition of the enhancement given in eq. 5 the random
  histograms $RR$ and $PP_R$ cancel in the process of obtaining the
  expression in eq. 6, so they are not used in the computation of our
  enhancement.

In the present work we perform two analyses: the cross-correlation of star
forming galaxies (as defined above) as primary galaxies with all galaxies as secondaries, and the
autocorrelation of star-forming galaxies. We will estimate the errors in
  our mark by means of bootstrapping resampling.

\subsection{Visual Morphologies} \label{sec:morph}

A particular challenge encountered when constructing a census of star
formation in pairs and mergers is accounting for systems with separations 
of
$<2"$ (which corresponds to $< 15$\, kpc, the radius within which we can no longer separate two massive 
galaxies
using COMBO-17; \citealp{bell06}). In order to pick
up the SF in all the stages of the interaction, we need to have an 
estimate of
the SFR not only in galaxy pairs with separations $>15$\,kpc but also in 
extremely close pairs and in recent merger remnants. 
We conduct our census of such close physical
pairs by including in the $<15$\,kpc range sources that are not resolved by 
COMBO-17, but appear to
be interacting pairs or merger remnants on the basis of visual 
classification
of the $\sim0.1\arcsec$ resolution ACS images. We try to recover
visually all $<15$\,kpc separation pairs of two $M_{*} > 10^{10}M_{\sun}$ 
galaxies with a mass ratio
between 1:1 and 1:4 missed by COMBO-17. In addition to those extremely 
close
pairs, we also account for the SF in recent merger remnants $M_{*} >2 
\times 10^{10}M_{\sun}$ (two times the minimum mass
of a galaxy in the sample and the minimum possible mass of a galaxy pair 
as
defined before).

\subsubsection{Discussion of visual classifications}

Our goal is to include very close pairs or already-coalesced major merger 
remnants into the census of `mergers' in order
to account for any SF triggered by the merger/interaction 
process\footnote{
Note that a consistent comparison with the projected correlation function
sample requires the inclusion of all non-interacting pairs that are
physically-associated (in the same cluster, filament, etc), are seen to be close projected pairs on the sky, 
but may be separated by as much as a few Mpc along the line of sight.
}.We do so on the basis of visual classification of the
sample.  The motivation for visual classification is a pragmatic one:
while a number of automated morphological classification systems have been 
developed in the last 15 years
\citep[i.e.][etc.]{abraham, cas, lotz}, 
it seems that the sensitivity of the observables
used (asymmetry, clumpiness, Gini coefficient, 
second order moment of the $20\%$
brightest pixels) is insufficient for matching the performance of 
visual classification in
current intermediate redshift galaxies with the same level of precision 
that
they display in the local Universe samples used for their calibration
\citep{cas, lisker, shardha}.

Yet, there is a degree of subjectivity to 
what one deems to be a major merger remnant.  Many factors
shape the morphology
of a galaxy merger that are beyond the control of the
classifier.  Bulge-to-total (B/T) mass ratios have an strong effect on both the
intensity of the SFR enhancement and the time at which the intensity peak
shall occur \citep[e.g.,][]{MH96}.
Orbital parameters strongly shape the development of easily
recognizable tidal tails and bridges (coplanar or not, retrograde
vs. prograde, etc). Prior dust and gas content of the 
parent galaxies (`dry' vs. `wet' mergers) will make
a difference to the appearance of the final object during the
coalescence. Furthermore, merging timescales will  depend on
whether the galaxies are undergoing a first passage or are in the final
stages of the merger. Finally, there is a degeneracy between all these
parameters and the relative masses of the galaxies undergoing the
interaction, which makes difficult in some cases to distinguish the morphological
signatures of a major merger from those of a minor merger.

Some of these factors (e.g. gas fraction, B/T ratios, etc) will also affect the enhancement of
the SFR during the interaction \citep[e.g.][]{dimatteo07, dimatteo08, cox08}.  
While there is considerable merger-to-merger scatter, 
encounters of two gas rich disk galaxies with parallel spins tend to 
develop, on
average, the strongest morphological features, but at the same time are 
more
likely to throw out large amounts of cold gas in tidal tails, preventing 
the
funneling of this gas to the central regions. Thus, samples selected to 
have the strongest morphological
features may have an average SFR enhancement different from the actual 
mean
enhancement\footnote{This bias might also be present in the case of studies
  looking for signs of interactions in the host galaxies of AGNs, attempting
  to assess whether the AGN activity is preceded by a merger.}.

One practical issue is that of passband choice and shifting.
We choose to classify the F606W images of the GEMS and STAGES
fields (in STAGES because that is the only available HST passband and 
in GEMS for consistency and because F606W has higher S/N that the 
F850LP data).  
This corresponds to rest-frame $\sim430(330)nm$ at redshift 0.4(0.8).
In previous papers \citep{wolf05,bell05,shardha}, 
we have assessed whether the morphological census
derived from GEMS/STAGES would change significantly 
if carried out data a factor of 5 deeper from the GOODS project
(testing sensitivity to surface brightness limits), or 
if carried out at F850LP (always rest-frame optical at these
redshifts). We found that the population does not show 
significantly different morphologies between our (comparatively)
shallow F606W data and the deeper/redder imaging data from GOODS
\citep[see Fig. 5 in][]{shardha}.

\subsubsection{Method}

\begin{figure*}
\begin{center}
\setlength{\unitlength}{1mm}
\begin{picture}(150,100)
\put(0,0){\includegraphics{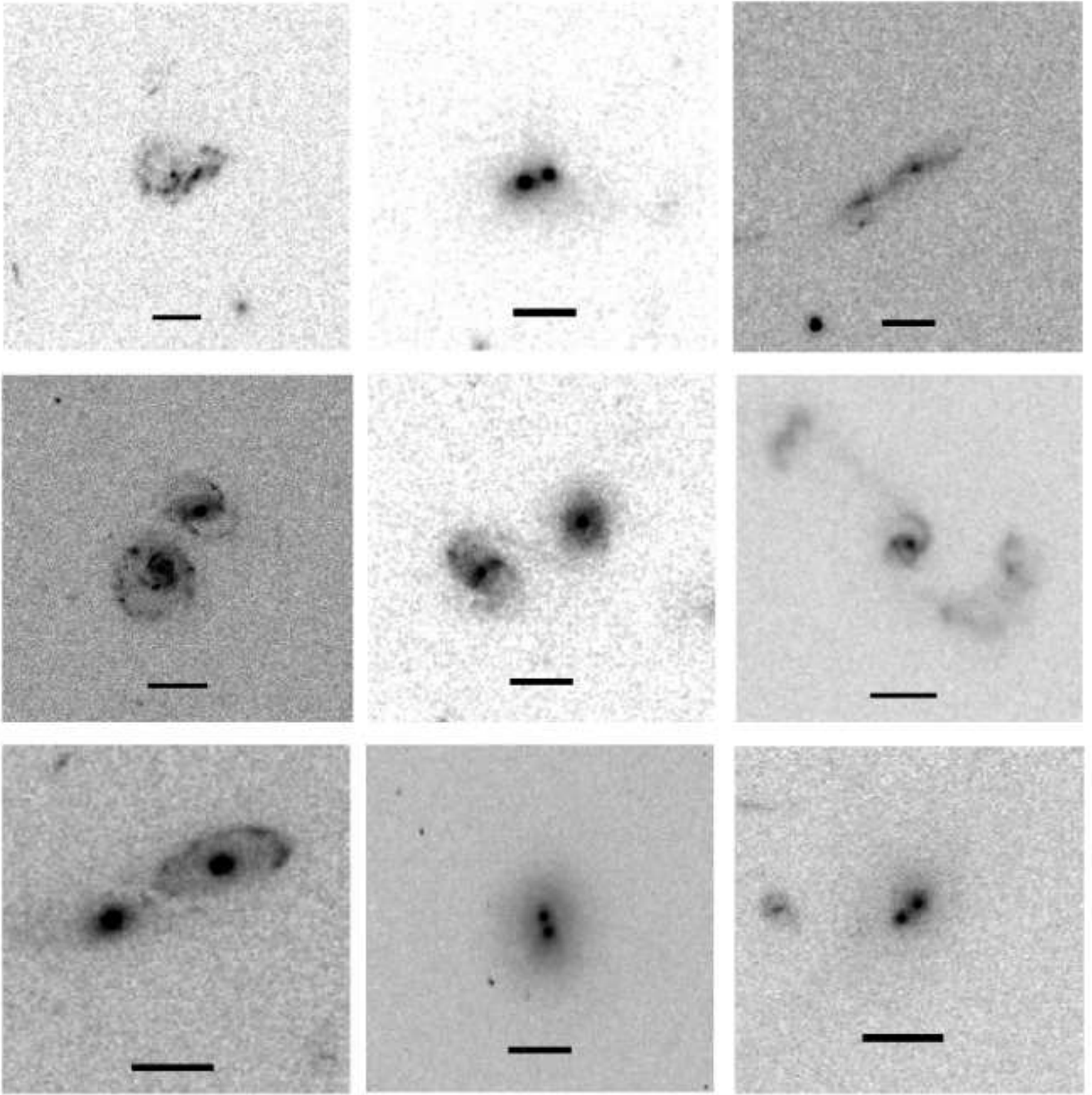}}
\end{picture}
\vspace{9.5cm}
\caption{\label{fig:major}
\emph{
Objects classified in group 2: Major close interactions. The presence of two
galaxies and signs of interaction are required. The classifier believes the
mass ratio is between 1:1 and 1:4. At this stage of the interaction, dry
mergers are still recognizable as seen in panels at top center, bottom center
and bottom right. The black bar at the bottom of every panel shows a proper distance of
20 kpc at the redshift of the object. Some of the objects classified in this
group were also separated as two galaxies in the ground-based catalog and treated in consequence.
}}
\end{center}
\end{figure*}

\begin{figure*}
\begin{center}
\setlength{\unitlength}{1mm}
\begin{picture}(150,100)
\put(0,0){\includegraphics{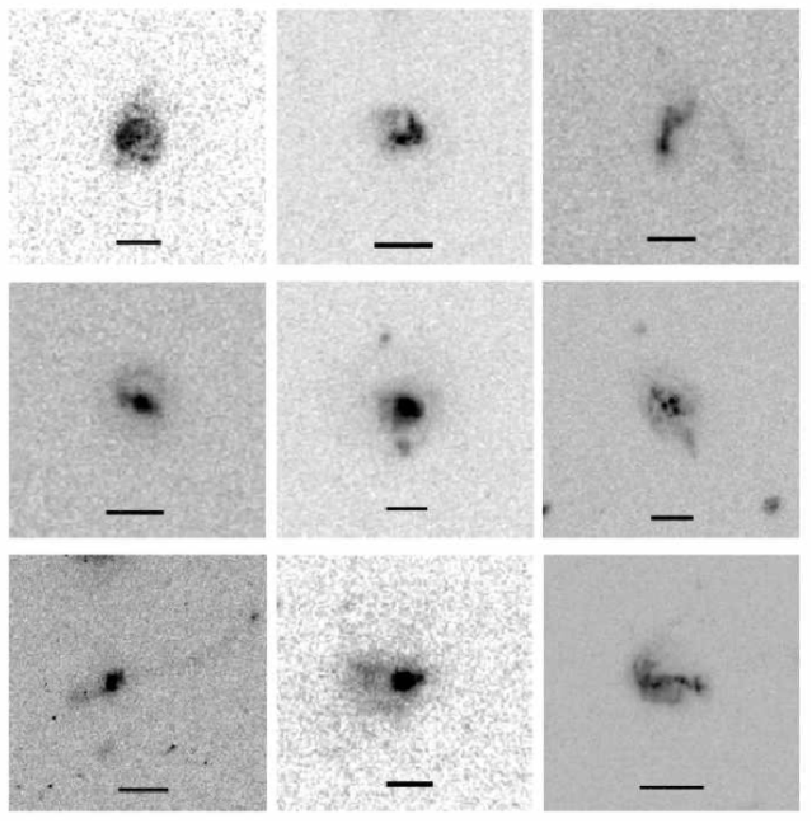}}
\end{picture}
\vspace{8.3cm}
\caption{\label{fig:remn}
\emph{
Objects classified in group 3: Major merger remnants.The black bar at the bottom of every panel shows a proper distance of
20 kpc at the redshift of the object
}}
\end{center}
\end{figure*}

An independent visual inspection of the galaxy sample has 
been carried out by four classifiers, A.R.R., E.F.B., R.E.S. 
and D.H.M. in order to
identify morphological signatures of major gravitational
interactions. Each classifier assigned every one of the 
$\sim$2500 sample members to one of the three following 
groups:
\begin{enumerate}
\item Non-major interactions: The bulk of galaxies in this bin show no 
signatures of gravitational interactions. Asymmetric, irregular galaxies 
with
patchy star formation triggered by internal processes 
lie in this category. A small fraction of galaxies in this bin show a clearly
recognizable morphology (e.g. spiral structure) but also 
signatures of an interaction (such as tidal tails, or warped, thick or 
lopsided disks) but have no clear interaction companion; note that these
objects could be interacting systems where the companion is now reasonably 
distant and/or faint and more difficult to identify. The tidal enhancement of SF from 
such systems will {\it not} be missed by putting them in this bin; rather, 
it will be measured statistically and robustly from the two point 
correlation function analysis.  Minor mergers and interactions 
(interactions
where the secondary is believed, on the basis of luminosity
ratio, to be less than 1/4 of the mass of the primary) also 
belong to this category.  

\item  Major close interactions: Close pairs resolved in HST imaging but
    not in ground-based COMBO-17 data, consisting
  of two galaxies with mass ratios between 1:1 and 1:4 based on relative
  luminosity, and clear signatures of tidal interaction such as tidal tails,
  bridges or common envelope (see Fig.~\ref{fig:major}). From now on we shall
refer to objects classified in this group as "very close pairs".

\item Major merger remnants: Objects that are believed to be the 
coalesced product of a recent major merger between two individual 
galaxies.   
Signposts of major 
merger remnants include a highly-disturbed 'train wreck' morphology, 
double nuclei of similar luminosity, tidal tails of similar length, 
or spheroidal remnants with large-scale
tidal debris (see Fig. ~\ref{fig:remn}).
Galaxies with clear signs of past merging but a prominent disk (e.g., 
highly asymmetric spiral arms or one tidal tail) were deemed
to be minor merger remnants and were assigned into the group 1.
Naturally there is some uncertainty and subjectivity
in the assignment of this class, in particular; such 
uncertainty is taken into account in our analysis by the 
monte Carlo sampling of all four classifications in order
to properly estimate the dispersion in the opinions of the 
individual classifiers (see below).
\end{enumerate}

\begin{table*}[h!]
\caption{Results from the morphological classification {\label{morph}}}
\begin{center}
\begin{tabular}{ccccc}
\hline
\hline
Lower mass limit & Sample size &  Group 1 &  Group 2  & Group 3 \\

\hline
$10^{10} M_\sun$ & 2551 & $2380 \pm 37 \pm 49$ &$106\pm 7 \pm 10$& $72\pm 7 \pm 8$ \\
$2\times 10^{10} M_\sun$ & 1749 &$1640 \pm 32 \pm 40$ &$69 \pm 6 \pm 8$  & $44 \pm 5 \pm 7$\\
\hline
\end{tabular}
\\
\vspace{-1cm}
\tablecomments{Galaxy and interaction sample. Group 1: Isolated objects and
  minor interactions. Group 2: Extremely close pairs ($r_P<15$\,kpc). Group 3:
  Merger remnants. The first error bar represents
  classifier-to-classifier scatter while the second one represents Poisson noise. 
\label{tab:mrgno}}
\end{center}
\end{table*}

We then assign 
the objects in the groups 2 and 3
(very close pairs with morphological signatures of interaction and merger
remnants respectively) to a small projected separation and treat every one 
of
them as a galaxy  pair in order to combine them with the correlation 
function
analysis result for pairs with separations $>2\arcsec$. All objects in Group 2
(extremely close pairs with projected separations < 15 kpc as measured by
centroids in HST imaging) are
assigned to a separation of 10 kpc and all objects in Group 3 (merger
remnants) are assigned to
a separation of 0 kpc. We have checked 
for duplicate pairs in both the
visually selected sample and the COMBO-17 catalog in order to avoid 
repeated
pairs. Galaxies in group 1 are already
included in the two point correlation function analysis, and any 
SF triggered by major interactions or early-stage major merging 
is accounted for by that method. As we 
have four different
classifications for every object (one given by each human classifier), we
randomly assign one of them,
calculate the average value of the weight we are using and repeat the 
process
a number of times. This approach presents two clear advantages: a) the 
resultant bootstrapping error 
not
only represents the statistical dispersion but also the different criteria 
of
the four human classifiers, and b) the morphology of every object is 
weighted
with {\it the four} classifications given. This means that objects with
discrepant classifications are not just assigned to one category when we
calculate the SFR (or specific SFR) enhancement; rather, 
any dispersion in classifications is naturally accounted for 
(e.g. minor/major criteria).  The numbers of such systems and their
uncertainties, estimated from the classifier-to-classifier scatter, 
are given in Table \ref{tab:mrgno}.

\section{Results} \label{results}

We are now in a position to quantify the triggering of star formation 
in galaxy interactions and mergers in the redshift interval 
$0.4<z<0.8$, in the cases where each galaxy has $M_* > 10^{10} M_{\sun}$
and the pair has a stellar mass ratio between 1:1 and 1:4.  
Our primary analysis is based on a marked cross-correlation between 
star-forming galaxies, as defined in \S3, and all galaxies in the sample. For morphologically-selected
very close pairs or interactions (unresolved by COMBO-17), we also require them to
be blue or detected by Spitzer to be considered as part of 
the star-forming sample\footnote{All galaxies, irrespective of
their color or IR flux, were classified; the star-forming galaxies
are simply a subsample of this larger sample.}, with a mass of $M_* > 2
\times 10^{10} M_{\sun}$.

We perform two analyses in this paper: the cross-correlation of star-formers as
primary galaxies with all galaxies as secondaries (our default case), and the
autocorrelation of star-forming galaxies.
While the first analysis is a rather more direct attack on the 
question of interest, we show results from the autocorrelation 
of star-forming galaxies to illustrate
the effects of making different sample choices on the final 
results.

\subsection{Enhancement in the Star Formation Activity}\label{enh}

\begin{figure*}[ht]

\begin{center}
\includegraphics[width=15cm,height=9cm]{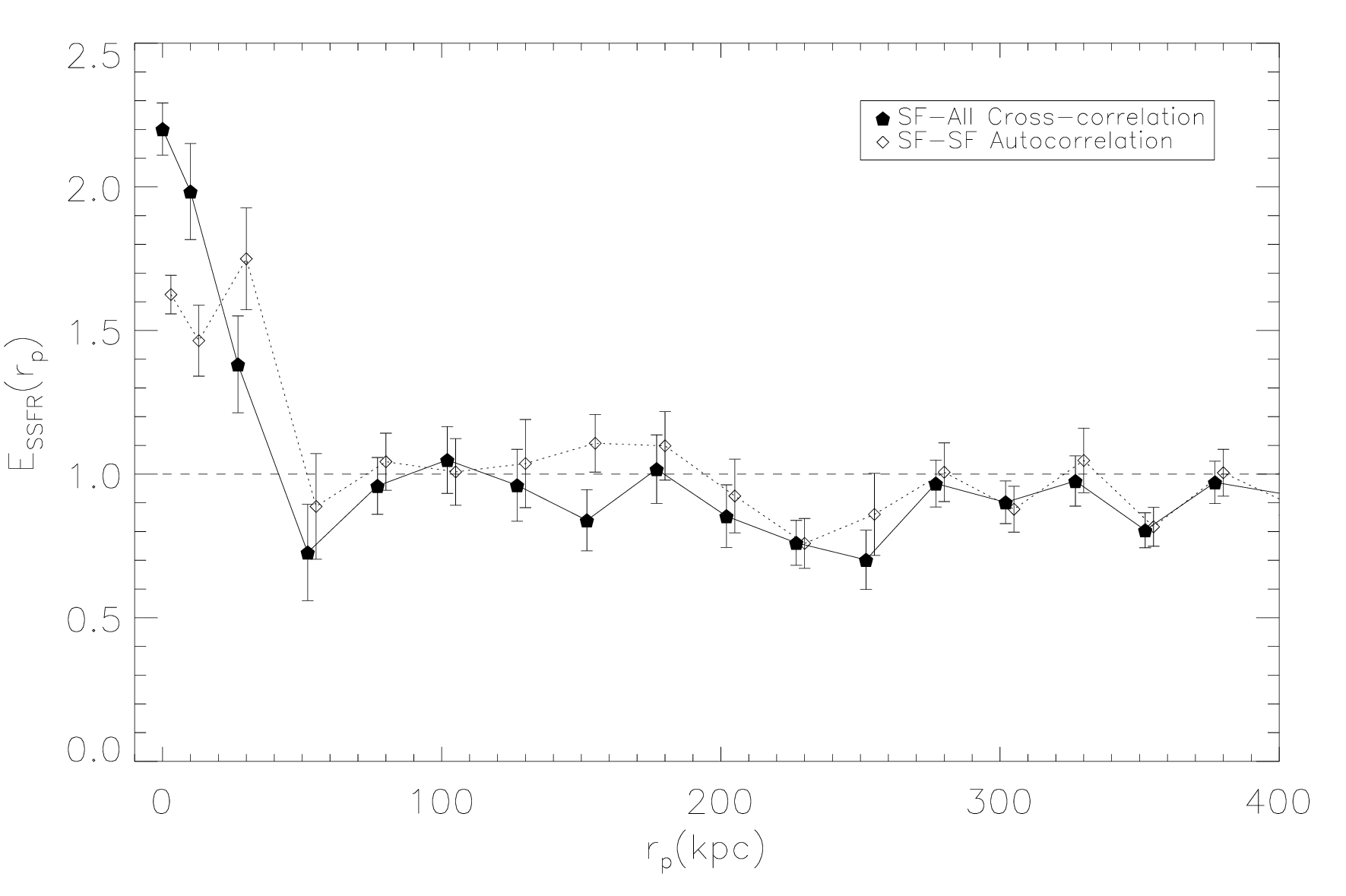}
\caption{ \label{fig:ssfr}
\emph{Pair specific SFR enhancement as function of the projected separation between two
galaxies.  The two smallest 
radii bins are derived from morphologically-selected very close 
pairs (shown with $r_P \sim 10$\,kpc) and merger remnants (shown with $r_P=0$); 
enhancements at larger radii are determined using weighted
two-point correlation functions. A statistically significant enhancement is present in galaxy pairs and
mergers below 40 kpc in both the cross--correlation between star forming
galaxies as primaries and all galaxies as secondaries (black filled symbols) and the autocorrelation
of star--forming galaxies (empty diamonds). Error bars have been calculated by bootstrapping.}
}
\end{center}
\end{figure*}

Our main results are shown in Fig.~\ref{fig:ssfr}, which shows the
enhancement of the specific star formation rate (SSFR) in pairs as a
function of their projected separation. As explained in \S\ref{mips} we
  use UV+IR SFRs for the objects detected in 24 \um and only UV SFRs for those
undetected. For the whole sample, 38\% of the galaxies
where detected by {\it Spitzer} above the 83\um limit, while if we restrict to
the groups 2 and 3 in our morphological classification we find a detected
fraction of 60\%. Fig.~\ref{fig:ssfr} shows a
clear enhancement in the SSFR for projected pair separations $r_P <
40$\,kpc. It could be argued that the SSFR is a better measure of the
SF enhancement than the SFR-weighted estimator, because the strong
scaling of SFR with galaxy mass is factored out. The figure shows both
the cross-correlation between star-forming primaries and all
secondaries (SF--All, solid line) and the star-forming galaxy
autocorrelation (SF--SF, dotted line).  The two bins at $r_p \le
15$\,kpc are calculated from morphologically-selected very close pairs
($r_p = 10$\,kpc) and merger remnants ($r_P=0$).  All the errors in
$E_{SSFR}$ have been computed by bootstrap resampling. This approach
allows us to treat both the morphologically-selected objects and the
galaxy pairs exactly in the same way, having as a result a coherent
display of the error bars.

There are two reasons why this excess in  $E_{SSFR}$ in close
pairs and remnants is likely a sign that interactions induce additional star
formation, rather than being due to a correlation with some other unidentified
quantity: a) It is well known
from simulations \citep{MH96, dimatteo07, cox08} that a burst of star
formation is expected in the collisions of gas-rich galaxies, and b)
the observed effect is in the opposite sense of the usual SFR--density
relation \citep[e.g.,][]{balogh}, which says that galaxies in dense
environments (where preferentially close galaxy pairs tend to be
found, as shown in \citealp{barton07}) have, on average, weaker star
formation activity than galaxies in less dense regions.

Even when we consider our morphological classification and further
  Monte Carlo resampling method to be very robust, potential classification
  errors could act in two different directions.
  Interacting systems misidentified as non-interacting will be diluted into the
  background star formation as single galaxies contributing to pairs at random
separations. While this SF should be lost to the interacting bin, the effect on the average SFR would be minimal. On the
other hand, isolated galaxies misidentified as interacting systems because of
internal instabilities or stochastic star formation would act to reduce the enhancement.

As mentioned before, the SFR for the objects undetected at
24 \um has been calculated based only on the UV. In the 24\um detected
objects, we have found no
clear trend in both the UV vs. UV+TIR SFRs and in the TIR/TUV vs. optical
dust attenuation but found instead a constant correction factor with a large
scatter ($4.1 \pm 2.4$ as estimated from the relation between TIR/TUV vs
  optical attenuation.) We have checked the effects of such a dust-correction of the UV-only SFRs:
the results differ in all bins by <10\%, comparable to or smaller than
other sources of systematic uncertainty.

\begin{figure*}[ht]
\begin{center}
\includegraphics[width=10cm,height=7cm]{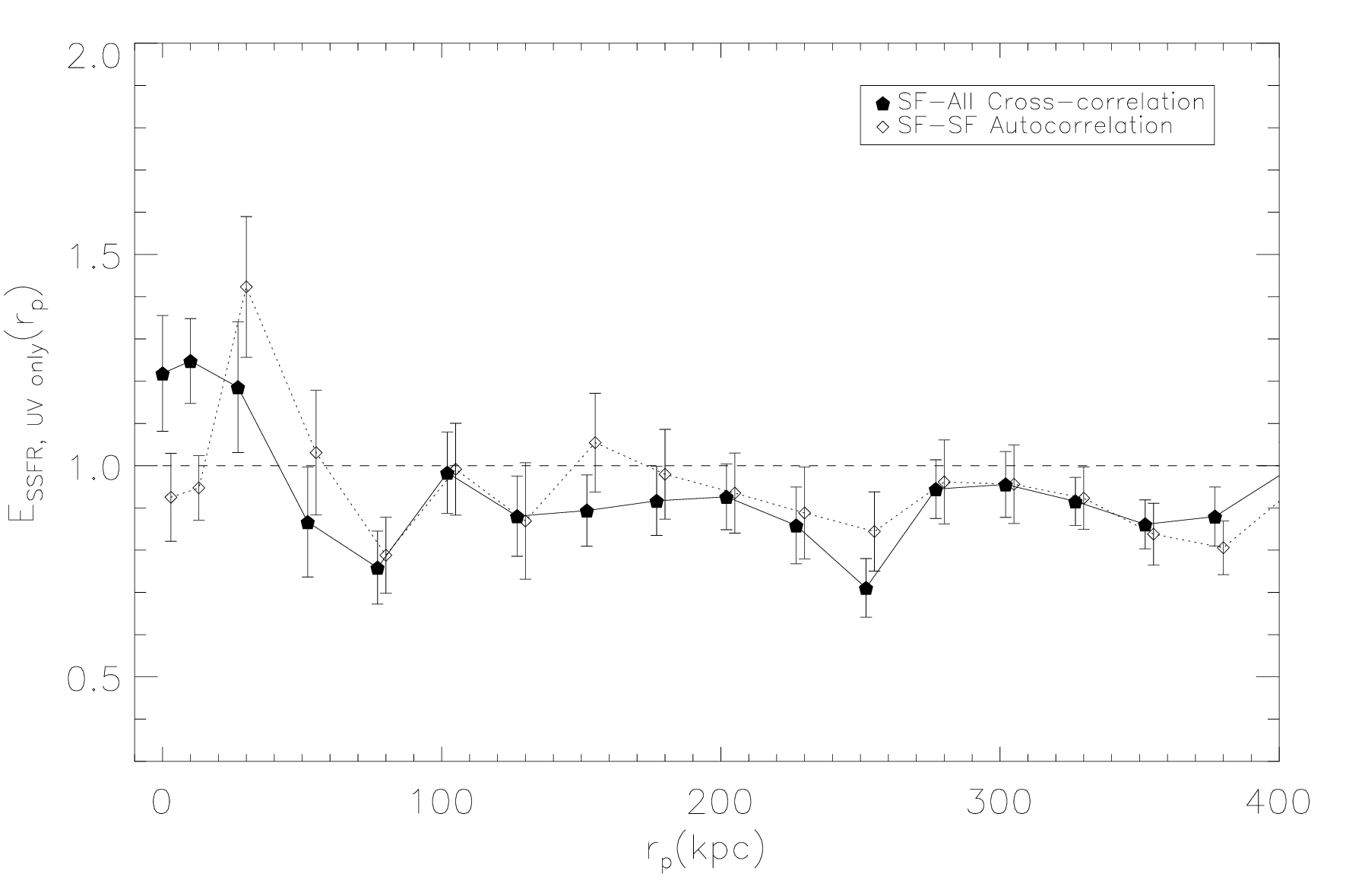}
\caption{ \label{fig:ssfr_uv}{\it Same as Fig.~\ref{fig:ssfr} but tracing only
  unobscured (UV-derived) star formation. The unobscured SSFR enhancement found in galaxy
  pairs with separations $r_P<40$\, kpc and mergers remnants is dramatically
  reduced with respect to the case in which the obscured star formation is taken
  into account (Fig.~\ref{fig:ssfr}).}
}

\end{center}
\end{figure*}

Yet, in order to understand the degree of obscuration in galaxy
interactions we have repeated our analysis including {\it only}
UV--derived SFRs, this is, excluding the TIR component in
Eq.~\ref{eqn:sfr} for 24\um detections. The result of this analysis is shown in
Fig.~\ref{fig:ssfr_uv}. The enhancement in the unobscured SSFR
measured for close pairs ($r_P<40$\,kpc) in this case is dramatically smaller than the enhancement
including the dust-obscured (IR-derived) star formation rate. This is
more apparent in the very close pairs and merger remnants, where the
excess in the SSFR even disappears completely in the case of the
SF--SF autocorrelation ($E(r_P<15{\rm kpc})\simeq1$). This implies
that most of the directly triggered star formation is dust obscured,
in good agreement with the expectations from \citet{mh94, MH96},
\citet{dimatteo07}, \citet{cox08} and the detailed models by
\citet{jonsson}. In these simulations most of the star formation is
triggered in the central regions of the galaxy after the cold gas has
been funneled to the inner kpc.

This scenario is also supported by our measurement of the mean ratio between
  the total SFR and the UV-derived, which gives an idea of the degree of
  dust-obscuration ($SFR_{IR+UV}/SFR_{UV}$). We find $6.64\pm0.66$ in the case of the
merger remnants (Group 3 in \S\ref{sec:morph}) and  $6.63\pm0.64$ in the case of
the very close pairs (group 2), compared to $3.15\pm0.53$  for all objects
in the sample.

\subsubsection{Star formation rate vs. specific star formation rate}\label{ssfr_sfr}

To study the fraction of the
  global star formation {\it directly triggered} by galaxy--galaxy
  interactions the enhancement in the SFR (rather than in the SSFR) is a
  better quantity to consider.

\begin{figure*}[h!]
\begin{center}
\includegraphics[width=10cm,height=7cm]{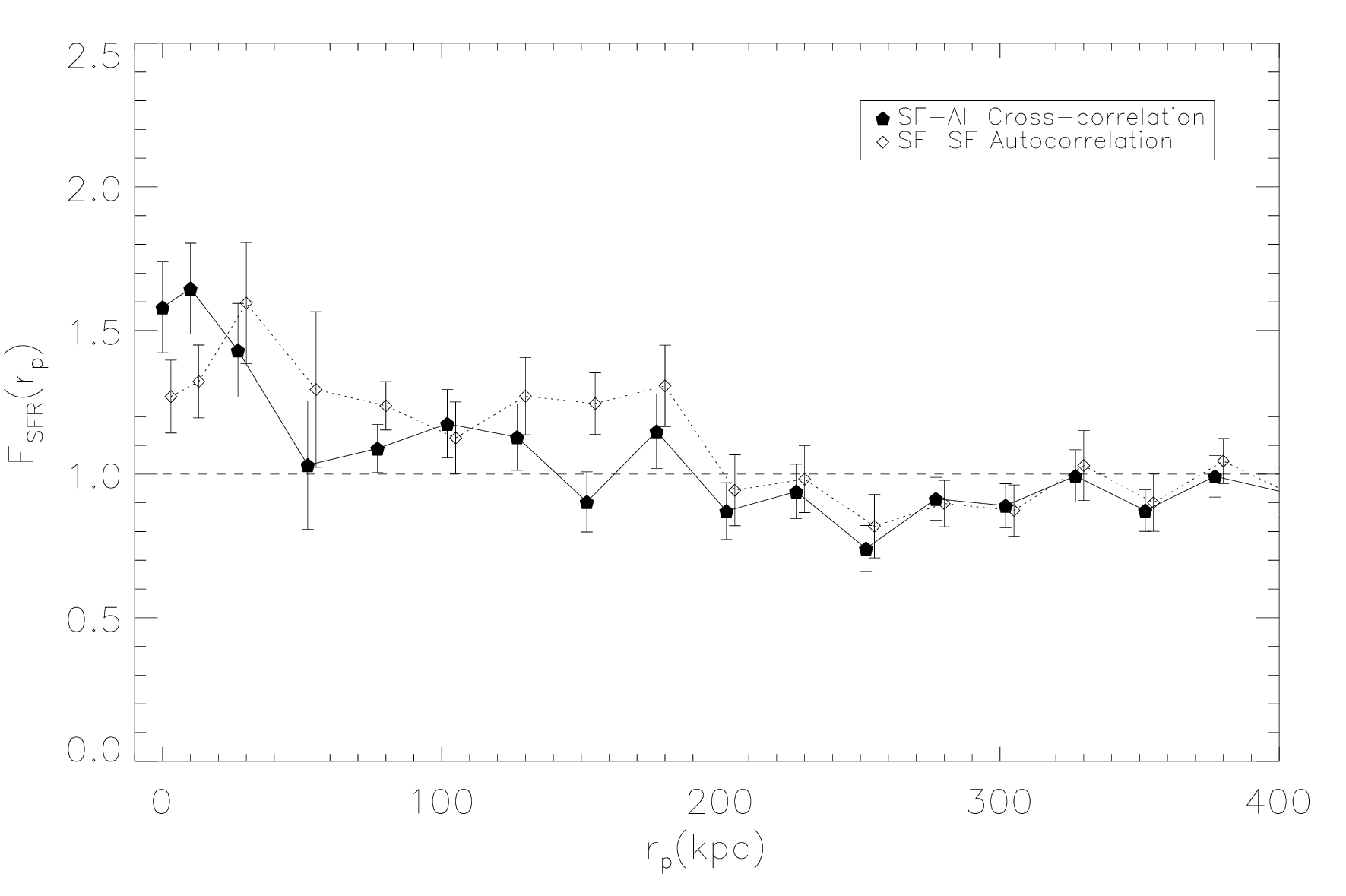}
\caption{ \label{fig:sfr}
\emph{SFR enhancement in galaxy interactions. The two smallest 
radii bins are derived from morphologically-selected extremely close 
pairs ($r_P \sim 10$\,kpc) and merger remnants (shown with $r_P=0$); 
enhancements at larger radii are determined using weighted
two-point correlation functions. There is a clear enhancement at $r_P<40$\,kpc
for the cross-correlation analysis (black-filled symbols) which is compatible
with $E_{SSFR}$ (Fig.~\ref{fig:ssfr}) except for the merger remnants, where
the excess is $\sim$50\% lower. The autocorrelation of star forming galaxies (empty
symbols) presents an unexpected behavior, showing a very mild enhancement at $r_P<180$\,kpc.
}}
\end{center}
\end{figure*}

We show in Fig.~\ref{fig:sfr} the enhancement in the SFR ($E_{SFR}(r_p)$) as
a function of the projected pair separation. For the cross--correlation function (our default case) the enhancement in
the SFR is similar to the one found in the SSFR at all separations except for
the merger remnants ($r_p=0$), in which the excess above the whole population is $\sim 50\%$ lower. The SFR--weighted autocorrelation of star forming
galaxies matches that  of the SSFR--weighted one for $r_p<40$\,kpc but differs
beyond: $E_{SFR}=1.25$ for $40<r_P<180$\, kpc. While most of these points in Fig.~\ref{fig:sfr}
are individually compatible with the error bars shown in Fig.~\ref{fig:ssfr},
 taken together they represent a $\sim 2\sigma$ significant difference
 between $E_{SSFR}$ and $E_{SFR}$ for the entire region $40<r_p<180$\,kpc.

\begin{figure*}[h!]
\begin{center}
\includegraphics[width=10cm,height=7cm]{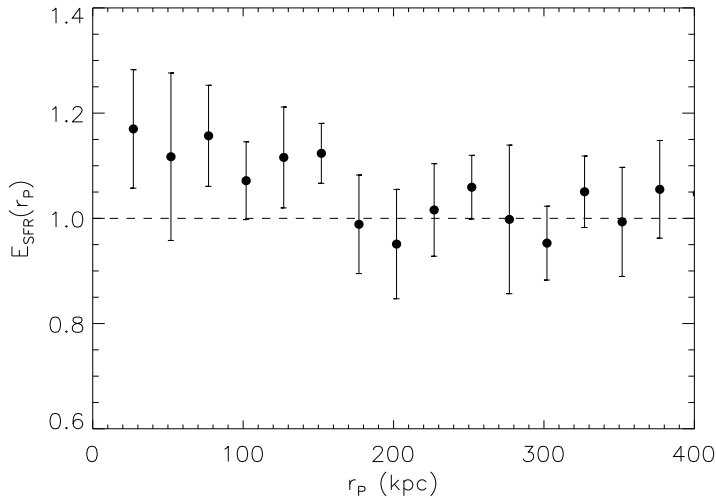}
\caption{ \label{fig:swap}{SFR enhancement measured after randomizing the SFR
    between galaxies of similar stellar mass. A mild enhancement is found out
    to separations of $\sim 160$\,kpc. We show the points corresponding to the
    SF-SF autocorrelation at distances $>15$\,kpc, where no morphological
    information is used.
}}
\end{center}
\end{figure*}

A potential driver of the SFR enhancement in the regime $40<r_P<180$\,kpc is
the fact that more massive galaxies tend to
be both more clustered and have higher SFR \citep{noeske}; this could translate into a weak
enhancement in the SFR in galaxy pairs living in dense environments
\citep[see][for a thorough discussion on the relation between galaxy pairs and
environment]{barton07} which will not
be present in the SSFR, because the normalization by galaxy mass factors out
this dependence. To test the relevance of this systematic effect, we randomized the SFRs among
galaxies of similar mass 500 times in the sample and repeated
the analysis. We show the results of this exercise in Fig.~\ref{fig:swap},
where we can see a tail of enhancement with a behavior similar to the
one seen in Fig.~\ref{fig:sfr}. We believe that a
combination of the density--mass--SFR relation plus noise is driving $E_{SFR}>1$
(autocorrelation) between 40 and 180\,kpc.

Accordingly, we consider only the enhancement at $r_P < 40$\,kpc as produced
by major merging in what 
follows, and use the differences between the SFR and SSFR enhancement on 
$<40$\,kpc scales as a measure of systematic uncertainty.  Under those 
assumptions, we find a weak enhancement of star formation at $r_P < 
40$\,kpc of $\epsilon = 1.50 \pm 0.25$ in the SF--SF autocorrelation and $\epsilon = 1.80
\pm 0.30$ in the SF--All cross-correlation. These values have been computed as
the average of the enhancement in the bins $r_P<40$\,kpc {\it together} in
$E_{SSFR}$ and $E_{SFR}$. These (conservative) error bars include both the
statistical uncertainties and the systematics driving the differences between
the SFR and the SSFR.

\subsubsection{Further Uncertainties}\label{uncertainties}
 As we have briefly mentioned in \S\ref{data}, there are some uncertainties
 which need to be estimated in the process of calculating the enhancement in
 the SF activity. Here we try to estimate the impact of the stellar-mass and
 IR SED selection uncertainties. Through this section we will focus in our
 default case, the SF-All cross-correlation.

\begin{figure}[h!]
\begin{center}
\includegraphics[width=10cm,height=7cm]{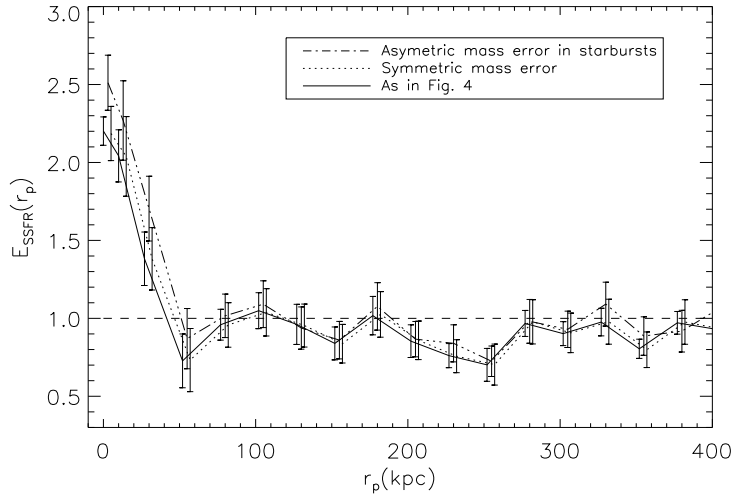}
\caption{\emph{ Enhancement in the SSFR including estimates for the errors in
    the stellar masses. Dotted line: Gaussian error with $\sigma = 0.1$ dex. Dash-dotted line: Non star forming galaxies with gaussian error with $\sigma
    = 0.1$ dex and starburst galaxies ($SFR>16M_{\sun}yr^{-1}$) with errors
    following an inverted lognormal distribution to produce a tail to the
    lower masses, with a shift of 0.1 dex also to
    the lower masses and $\sigma = 0.2$ dex. Solid line: Enhancement in the
    SSFR as in Fig.\ref{fig:ssfr}, for comparison.
\label{fig:mass_errors}
}}
\end{center}
\end{figure}

 Random errors in stellar masses in \cite{borch} are $< 0.1$ dex (with 0.3
   dex in cases with large starbursts \citep{bell01}) on a
 galaxy-by-galaxy basis, and systematic errors non related to the choice
   of an universally-applied stellar IMF are 0.1 dex. In
 addition, M/L ratios in starbursting galaxies can be biased to produce
 unrealistic high stellar masses\citep{bell01}. Those effects would have certain impact in
 the calculation of the SSFR, and thus, in the enhancement of that
 quantity. In order to estimate how those mass uncertainties affect our results, we
 have run two additional Monte Carlo shufflings. In Fig.~\ref{fig:mass_errors}
 we show the result of this exercise. We have randomly added a gaussian error with
 $\sigma=0.1$ dex to the stellar masses of all galaxies and repeated the
 process 500 times, finding an average output value similar to the one
 presented in Fig.~\ref{fig:ssfr} but with larger errors. The impact of the new
 errors on the average enhancement (taking into account also the enhancement
 in the SFR, as we did in the previous section) is negligible. 

 In order to estimate the uncertainties introduced by systematics in the M/L
 ratio of starbursts, we have performed a similar exercise but using an error
 which includes a systematic shift down of 0.1 dex, $\sigma=0.2$ dex and a
 tail to the lower masses defined by an inverted lognormal distribution. We
 have applied this new error to objects with $SFR>16M_{\sun} yr^{-1}$, which roughly
 corresponds to twice the average SFR in our sample of {\it star forming}
 galaxies, and the symmetric error described above to galaxies with
 $SFR<16M_{\sun} yr^{-1}$. The
 result (dash-dotted in Fig.~\ref{fig:mass_errors}) shows some extra enhancement
 in this case, which leads to an average enhancement in the SF activity
 $\epsilon=1.85 \pm 0.35$, barely changing the result already found.

 Another potential source of uncertainty is the stellar masses of pairs of
 galaxies not resolved in the ground-based photometry catalog (i.e., our very
 close pairs group). In order to test the impact of this underdeblending on
 the galaxy masses, we take U and V rest-frame fluxes of galaxies widely
 separated, add them together and check what stellar mass would result in the
 case of applying the \cite{borch} method to a galaxy with exactly the same
 color as the combination of the two galaxies, and compare with the sum of
 the two original masses. We find that for pairs of galaxies of all kinds
 (all-all, SF-SF and SF-all) there is a $<0.01$ dex offset and 0.08 dex
 scatter between the two sets of masses. I.e., masses from combined
 luminosities are the same as the sum of the individual masses to 0.08 dex,
 what means that our stellar masses are extremely robust against
 underdeblending issues.

 Together with the stellar-masses, the main source of uncertainty in our
 analysis is the conversion between observed 24\um and TIR in the process of
 obtaining the SFRs. \cite{zheng07a}
 have demonstrated that the Sbc template used here is an appropriate choice for
 this dataset at all IR luminosities, but in order to find an absolute upper
 limit for the final results we will show in \S\ref{ss:trigger} we estimate
 the different results we would obtain if considering an Arp220 template in some cases.

\begin{figure}[h!]
\begin{center}
\includegraphics[width=10cm,height=7cm]{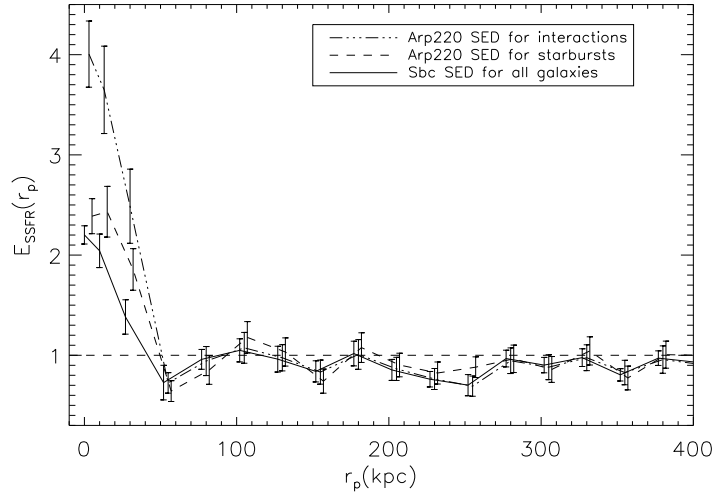}
\caption{\emph{ Impact on the SSFR enhancement when using an Arp220 template in the conversion
    between observed 24 \um and TIR luminosity for certain objects. Red line:
    Extreme case in which we apply an Arp220 template to all interacting
    systems (and Sbc template to everything else). Green line: Arp220 template
    applied to objects with $SFR>16 M_{\sun}yr^{-1}$ (and Sbc template to everything else). Black line: Same as in
    Fig. \ref{fig:ssfr}, for comparison. Red and green lines include the
    asymmetric stellar mass errors applied in Fig.~\ref{fig:mass_errors}.
\label{fig:sfr_errors}
}}
\end{center}
\end{figure}

 We find that at a given 24\um flux, the use of an Arp220 template
 gives a TIR luminosity which is higher than that derived using a Sbc template by a factor of
 2. We apply this factor of 2 correction to the TIR luminosity of all
 the galaxies that we define as starburst for this purpose
 ($SFR>16M_{\sun}yr^{-1}$) and show the result as the green line in
 Fig. \ref{fig:sfr_errors}. The enhancement found in this case is
 $\epsilon=2.1\pm 0.4$, consistent with, by higher than the $\epsilon=1.8\pm 0.3$ found in \S\ref{ssfr_sfr}.

We also want to test the extreme case in which the IR SED of all galaxies
undergoing an interaction follows an Arp220 SED, independently of their level
of SFR. This is clearly an unrealistic case as we know that some of our
galaxies in close pairs and remnants have SFRs as low as 4-5$M_{\sun}yr^{-1}$
(factor of 10 less SFR than Arp220), and we know also that the average IR SED
of $z\sim 0.6$ galaxies with $SFRs \ge 10 M_\sun yr^{-1}$ is similar to the
Sbc template adopted here \citep{zheng07b},  
but it is useful in the sense that it provides a strong upper limit beyond the
uncertainties of our data and method. The result found in this case can be
seen as the red line in Fig.~\ref{fig:sfr_errors}. Clearly, a much stronger
enhancement is present as a consequence of this overestimation of the TIR
luminosity, that leads, when taken together with the SFR enhancement
calculated in the same way, to $\epsilon=3.1\pm 0.6$.

\subsection{How important are mergers in triggering dust-obscured starbursts?}

We have demonstrated that when averaged over all events and all event phases there is a relatively modest SFR enhancement
from major galaxy merging and interactions. It is of interest to constrain
how the distribution of SFRs differs between the 
non-interacting and interacting galaxies. Here we present a preliminary result
on one aspect of the issue, namely the fraction of infrared-luminous galaxies that 
are in close pairs $r_p < 40$\,kpc or were visually classified
as merging systems.

\begin{figure}[ht!]
\begin{center}
\includegraphics[width=9cm,height=6cm]{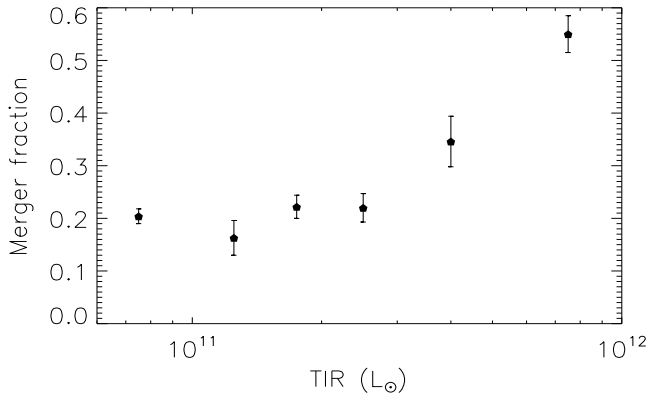}
\caption{\emph{The fraction of systems in close 
projected pairs $r_p < 40$\,kpc or in visually identified mergers as 
function of the total IR luminosity of all 24\um detected
  galaxies in the sample. The merger fraction is $\sim20\%$
  from our luminosity limit of $6 \times 10^{10} L_\sun$ to
   $3\times 10^{11} L_\sun$.  Higher 
  than this luminosity, the merger fraction begins to grow to $55\%$
  just below $10^{12}L_\sun$.  \label{fig:irfrac}
}}
\end{center}
\end{figure}

In Fig.\ ~\ref{fig:irfrac} we show how the fraction of 
galaxies that are either in close pairs ($r_p < 40$\,kpc) or in 
morphologically-classified merger remnants varies as a function of their total
IR luminosity.  This fraction is constant ($\sim$20\%) for $6 \times 10^{10}L_{\sun} < L_{TIR} < 3 \times
10^{11} L_{\sun}$. At higher luminosities, the merger fraction increases as a
function of the IR luminosity, reaching
 $55\%$ just below
$L_{TIR}=10^{12} L_{\sun}$. The lower IR limit of $6 \times
10^{10} L_\sun$ was chosen to ensure a flux of 83$\mu$Jy over the entire
redshift range.
The increase in merger fraction at high IR luminosity is in 
accord with previous results at both low and intermediate redshift \citep[e.g.,][]{sanders88}. This suggests that 
merging and interactions are an important trigger of intense, dust-obscured
star formation. Apparently, high IR luminosities are difficult to reach 
without an interaction.  

A key point, however, is that {\it not all} mergers have high IR
luminosity. While mergers can produce enormous SFRs, and also
$L_{TIR}>10^{12}L_{\sun}$ is best reached by merging, the typical SFR
enhancement in mergers is modest.

\section{Discussion} \label{disc}

We have assembled a unique data set for galaxies  at $0.4<z<0.8$ that combines
redshifts, stellar masses, SFRs and HST morphologies to explore the role of
major mergers and interactions in boosting the SFR. In practice, we have
combined projected correlation-function and
morphological techniques to estimate the average enhancement of
star formation in star-forming galaxies with $M_* > 10^{10} M_{\sun}$
and $0.4<z<0.8$, where the average is taken over most merging phases and all
mergers.  We find a SF enhancement by a modest factor of $\sim1.8$ for
separations of $< 40$ kpc in both the SFR
and the SSFR. How does this compare with 
previous observations and models? What implications does this mild enhancement
have on the contribution of major mergers to the cosmic SF history?

\subsection{Comparison with previous observations} \label{compar}

 \begin{figure*}[ht]

\begin{center}
\includegraphics[width=10cm,height=7cm]{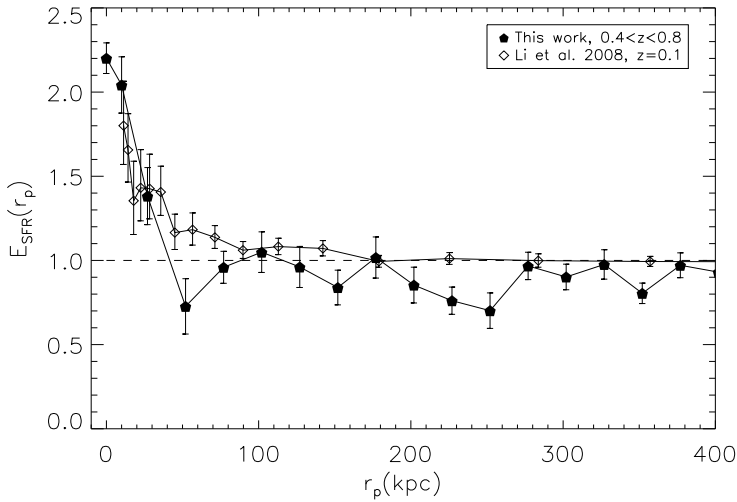}
\caption{ \label{fig:comp}
\emph{ Comparison between the enhancement found in this work (black filled points) and
  the one found in \cite{li} at $z\simeq 0.1$ (open diamonds).  In both cases, a 
  cross-correlation
  SF--all is shown. Both works show a statistically significant enhancement of the SSFR at 
  $r_P < 40\,$kpc.  
}}
\end{center}
\end{figure*}

Our analysis is most directly comparable to estimates of star formation rate enhancement in
galaxy close pairs by \cite{li} using the SDSS at $z \sim 0.1$ 
because of the similarities between our methods. 
Using a cross-correlation between star-forming and all galaxies, 
they found an enhancement of $\simeq 1.45$ for an average
galaxy mass of 
$\avg{log(M_*/M_\sun)}=10.6$ within a radius of 15 to $\sim 35-40$ kpc. 
In Fig.\ ~\ref{fig:comp}, we show the comparison between their present-epoch
measurements and ours (average galaxy mass $\sim 10^{10.5} M_{\sun}$, SF--all
cross-correlation) revealing reasonable quantitative agreement. Both the
projected separation scale ($\lesssim 40$\,kpc) and the overall amplitude at
small projections ($\times 1.5-2$) agree. The enhancement found here also agrees
(given the error bars) with the enhancement found at $0.75 < z < 1.1$ in
\cite{lin}. Our results are similar to those at both $z=0.1$ and at $z=1$, despite the factors of several difference
between the typical star formation rates of galaxies between 
$z=1$ and $z=0$ \citep[see, e.g.,][]{zheng07b}. This is interesting, 
and points to a picture in which at least the average 
enhancement of star formation in galaxy interactions
appears to be independent of the `pre-existing' star formation
in the population.

\cite{lin} also measured an enhancement in the TIR emission in galaxy pairs
and mergers in the $0.4<z<0.75$ range. They find that 
the infrared luminosity of close pairs with both members selected to be blue is 
$1.8\pm0.4$ times that of control pairs, similar to our value $1.75\pm 0.18 $
for close pairs in the bin $15kpc < r_P < 40kpc$ from the SF-SF autocorrelation. For late-phase mergers, they measure $2.1 \pm 0.4$,
marginally consistent with our $1.54\pm0.08$ from the SF-SF
autocorrelation at $r_P=0$\,kpc. Slight differences in that number may be attributed to
differences in the 'merger' classification. For example, many of the remnants that we
include in this study may not be detected by automated methods based on the
intensity-weighted Gini-M$_{20}$ or asymmetry parameters. As shown in recent
work \citep{shardha,miller}, automated methods based on CAS asymmetry
parameters tend to capture only a fraction (typically 50\% to 70\%) of the
visually-identified merger remnants and often pick up a dominant number of
non-interacting galaxies that have small-scale asymmetries associated with
dust and star formation.

In related work \citep{shardha}, we recently estimated the overall merging rate
and also addressed the SFR enhancement at $0.24<z<0.8$. For the subsample of
systems with $M_*>2.5\times10^{10}M_{\sun}$ we find that the average SFR of
late stage mergers with mass ratio between 1:1 and 1:10 (both major and minor
mergers) are only enhanced by a modest factor (1.5-2 from their Fig. 15) with respect to non-interacting
galaxies. There are three differences that make it difficult to perform an exact
comparison between our works: a) In the present study we try to isolate the
contribution from major interactions (mass ratio 1:1 to 1:4) while in
\citet{shardha} we focused on both major and minor interactions (mass ratio from 1:1 to 1:10).
b) The normalization is slightly different because in the present work we
compare the SFR in mergers with the SFR in the pair of progenitors, while in
\citet{shardha} we compare with the SFR of individual galaxies with mass similar to the
interacting system. As the average SFR is a function of the galaxy stellar
mass, $2\times$SFR$_{M>10^{10},progenitor}\neq\,$SFR$_{M>2\times10^{10},descendant}$.
c) In the present paper we attempt to target both early and late phases of the
interaction, while in \citet{shardha} we focus on the later
phase. Nonetheless, it is encouraging that the two studies agree qualitatively
in finding a modest enhancement in the average SFR in galaxy interactions.

Taken together, we argue that our results are consistent with 
those of previous works. We have used a bigger sample of galaxies with both
HST/ACS and {\it Spitzer}/MIPS coverage than previous works at $z\geq0.4$ and we tried to
trace the SFR enhancement in all the stages of the interaction with a
consistent treatment of ground-based selected galaxy pairs and morphologically-selected pairs and remnants. We view it as extremely encouraging 
that where the works are the most robust (close pairs), 
the results are highly consistent (comparing our work with 
\citealp{li} and the pairs from \citealp{lin}). It is clear that robustly
assessing the star formation enhancement in advanced-stage 
mergers, identifiable using only high-resolution data and morphological
techniques, is considerably more challenging.  The results 
for advanced-stage mergers are therefore less well-constrained, but 
are nonetheless all consistent with a modest but significant enhancement 
in SFR.

\subsection{What fraction of star formation is triggered by major interactions?}
\label{ss:trigger}

We can now combine our estimates for the SFR enhancement, the fraction of galaxies in projected close pairs, the 
average SFR, and the amount of SFR in recognizable merger remnants to quantify
what fraction of star formation at $0.4<z<0.8$ is {\it directly triggered} by
major interactions. We will not include systematics such as the uncertainty in 
conversion of 24{\micron} to total IR, or the effect of the 24{\micron}
flux limit, but we will consider the systematics driving the difference
between the SSFR enhancement and the SFR enhancement. We make the
cross-correlation between star forming galaxies as primaries and all galaxies
as secondaries our default case because it includes the residual SFR in red
galaxies and also traces the SF enhancement in disk galaxies during the
encounter with a non star forming galaxy. We will also show the values
obtained for the SF-SF autocorrelation. We use the values $1.80\pm0.30$ and
$1.50\pm0.25$ found in \S~\ref{ssfr_sfr} in  $r_P<40$\,kpc systems, for the cross-correlation and the autocorrelation respectively. 

The fraction of galaxies in close physical pairs within a separation $r_f$ can be derived using the 
following approximation 
\citep{patton00,masjedi06,bell06}: 
\begin{equation}\label{eq:fracpair}
P(r<r_f) = \frac{4 \pi n}{3 - \gamma} r_0^{\gamma} r_f^{3-\gamma}.
\label{cp}
\end{equation}
Here $P(r<r_f)$ is the fraction of galaxies in the parent sample in pairs with real 
separations of $r<r_f$, $n$ is the number density
of galaxies satisfying the pair selection criteria, and $r_0$ and $\gamma$ 
are the parameters of the power-law real-space correlation function of 
the parent sample, subjected to the pair selection criteria (i.e., we
use a stellar mass ratio of 1:1 to 1:4 as a requirement for a 
pair to enter into the correlation function).

Note that because in this paper we typically impose criteria for matching
and forming pairs (e.g., a mass ratio between 1:1 and 4:1), the number
density $n$ used is not the number density of the larger parent sample
$n_{\rm parent}$. The number of possible pairs at any projected separation
range is lower than in the case in which no mass ratio criteria is imposed
because many pairs with mass ratios beyond the allowed limit are automatically
rejected. As the fraction of galaxies in close physical pairs is directly
related to the number density of galaxies $n$, this parameter has to be
fine-tuned in order to get the right fraction. The number density used in
Eq.~\ref{eq:fracpair} has to be corrected for the effect that the mass ratio criteria
introduces on the total number of potential pairs. Then, the number density of
the larger parent sample $n_{\rm parent}$ is not used here, instead we use $n = n_{\rm parent} N_{\rm pairs}/[{0.5 N_{\rm parent} (N_{\rm parent}-1)}]$, 
where $N_{\rm pairs}$ is the number of pairs that can be formed in the parent
sample given the matching criteria, and $N_{\rm parent}$ is the number of galaxies in the parent sample ($N (N-1)/2$ is the expression for the number of possible pairs in the case of simply pairing up the parent sample).

We tested this approximation using the 
semi-analytic galaxy catalog of \cite{delucia06}, derived
from the Millennium $N$-body simulation.  At $z \sim 0.6$ 
these simulations matched well the stellar mass function and 
correlation function of $M_* > 10^{10} M_{\sun}$ galaxies.
We find that at $r < 50$\,kpc Equation \ref{cp} is a good approximation 
to the actual fraction of galaxies in close pairs in the simulation; at larger separation 
Equation \ref{cp} is increasingly incorrect (this is the subject of 
a paper in preparation).

From fits to the projected two-point cross-correlation function of our sample, 
we determine $r_0 = 1.8 \pm 0.2$\,Mpc, $\gamma = 2.2 \pm 0.1$ for the
real-space correlation function, and 
$n = 0.0152$\,galaxies per cubic Mpc\footnote{The correlation 
function is calculated in proper coordinates, because the process
of interest is galaxy merging and close pairs of galaxies
have completely decoupled from the Hubble flow.}; the latter gives
a sample of $N_{\rm gal} = nV = 1913$ galaxies in the 
volume probed by this study. 
 This yields
$P(r<40\,{\rm kpc}) = 0.06\pm0.01$ (i.e., 6\% of sample galaxies are in close
pairs with real-space separations $<40$\,kpc).  With this 
real-space two-point correlation function, 75$\pm10$\% of all projected 
close pairs should be real close physical pairs\footnote{ This is only valid
  after removing the effect introduced by
  purely random projections with the correlation function method.}
(Eq. 6 of \citealp{bell06} and confirmed using the Millennium Simulation at the
redshift of interest). Thus the fraction of objects
in projected close pairs will be $f_{\rm pair,proj} =
0.06/0.75=0.08\pm0.02$. This fraction includes projections due to real
structures like clusters or filaments, but not the purely random projections
due to redshift uncertainties
which would be present if we would just count the galaxies in projected
close pairs in our catalogue.

When considering {\it all pairs} at all separations in our sample with $M_*>10^{10} M_\sun$, mass
ratios between $1:1$ and $1:4$, and primary galaxies with 24{\micron}
fluxes $>83 \mu$ Jy and/or blue, the average SFR is $\left< SFR \right>_{\rm typical,pair} = 13.2 \pm 0.6 M_\sun{\rm yr^{-1}}$.
The {\it total} SFR in the $N_{\rm rem} = 38\pm 5$ recognizable merger remnants is 
$SFR_{\rm remnants}=753 \pm 97 M_\sun{\rm yr^{-1}}$. Thus, we can calculate the fraction of  
SFR occuring in pairs with separations $<40$\,kpc: 
\begin{equation}\label{eq:totfrac}
\frac{N_{\rm gal} f_{\rm pair,proj} 0.5 \epsilon \left<SFR \right>_{\rm typical,pair} +
  SFR_{\rm remnants}}{N_{\rm gal} 0.5 \left<SFR \right>_{\rm typical,pair}} = 20\pm3\%\
\end{equation}

\noindent for the SF--All correlation. A similar analysis with the SF--SF correlation
yields $16\pm3\%$. What we have done in the numerator of Eq.~\ref{eq:totfrac} is to take the
typical SFR in our pairs and divide by two in order to get the typical SFR of
a galaxy contributing to such pairs. This number is different from the typical
SFR in our galaxy sample for two reasons: first, we have imposed a mass--ratio criterion (only
allow pairs with mass ratios between 1:1 and 1:4), which makes the averaged SFR in all pairs to
be slightly biased high respect random pairs without any mass ratio criterion,
and second, the fact that we force the primary galaxy to be a star former (in
the case of the cross-correlation) has a similar effect. Then we have multiplied it by the
enhancement $\epsilon$ in order to take into account the excess SFR triggered
by major interactions and introduced the factor $N_{gal}$ to account for all
the SFR occuring in those galaxies. A key piece of Eq.~\ref{eq:totfrac} is the different
treatment of merger remnants. The correlation function can tell us what is the
fraction with separations between $r_P=40$\,kpc and $r_P=0$\,kpc but we have
defined merger remnants as objects which have {\it already} coalesced, so
if we think in terms of the duration of the interaction instead of the
separation between the galaxies, these objects would be {\it beyond} the reach
of the correlation function, and have to be treated separately. In the
denominator we have only divided by the total SFR occuring in {\it all} the
galaxies contributing to any pair we can form with the already mentioned
criteria. The difference between this factor and the total SFR calculated
simply adding up the SFR of all galaxies in the sample is 5\% and is a
consequence of the few galaxies which are not paired with any other galaxy in
the sample.

Yet, the fraction of the total SFR that occurs in pairs and remnants with $<40$\,kpc
separation does not immediately characterize the SFR {\it triggered} by
interactions, because $\sim 12\%$ of SF should happen at $r_P<40$\,kpc anyway,
as we show below. Only the {\it excess} star formation in pairs and remnants
 should be atributed to triggering by interactions:
\begin{equation}\label{eq:final}
\frac{N_{\rm gal} f_{\rm pair,proj} 0.5 (\epsilon-1) \left< SFR \right>_{\rm typical,pair} +
  (SFR_{\rm remnants}-N_{rem} \left< SFR \right>_{\rm typical,pair})}{N_{\rm gal} 0.5 \left< SFR \right>_{\rm typical,pair}} = 8\pm3\%.
\end{equation}
Again, a similar analysis for the SF--SF autocorrelation yields
$5\pm3\%$. These values for the excess are 12\% lower than those in
Eq.~\ref{eq:totfrac} due to the total number of interacting systems, which is
higher than the 8\% of galaxies in close pairs mentioned before because it includes the merger remnants
that are not taken into account by the correlation function method.

 Taking all this together, this analysis shows that only $\sim 8\%$ of the star formation at 
$0.4<z<0.8$ is triggered by major mergers/interactions. This may seem in
disagreement with previous  results from 'morphological' studies. We therefore
compare our results with those of \cite{bell05} and
\cite{wolf05}, who found that $\sim 30\%$ 
of the global SFR at $z=0.7$ is taking place in
morphologically perturbed systems and with \citet{shardha}, where we find a
similar result at $0.24<z<0.8$.

Both \cite{bell05} and \cite{wolf05} performed a study of the total SFR
occuring in visually-classified interacting galaxies in a thin redshift slice
$0.65<z<0.75$ without imposing a lower mass limit (only an apparent magnitude limit). That is the key difference
between those earlier works and ours. We impose a mass cut in this paper of
$10^{10}M_\sun$ and $2\times 10^{10}M_\sun$ for galaxies and visually-classified interactions respectively.
E.g. the fraction of SFR in galaxies with $M_* > 1-2\times 10^{10} M_{\sun}$ that \cite{bell05} and 
\cite{wolf05} identified as interacting/peculiar is 15-21\%, compared to
20\% in Eq.~\ref{eq:totfrac}. Only 1/6 of the 
star formation in the interacting/peculiar galaxies from \cite{bell05} 
and \cite{wolf05} occurs in what we would designate as merger remnants, with the other 5/6 occuring in 
galaxy pairs. Jogee et al. (2009) argued that 30\% of star formation was in systems that 
they classified as major or minor interactions, with mass limits different 
from those used in this paper. This value is an upper limit to the 
fraction of star formation in major mergers where each galaxy has mass 
$> 10^{10} M_{\sun}$, both because of the effect of mass limits, and 
because minor mergers host much of the star formation in systems that they
classified to be interacting.  Accounting for these differences, our 
result is in qualitative agreement with theirs.

However, the key difference is that neither \cite{bell05}, \cite{wolf05} nor
\citet{shardha} try to quantify
the {\it excess} of SF in interacting systems, as we do in going from
Eq.~\ref{eq:totfrac} to Eq.~\ref{eq:final}. In summary, our new results here
pose no inconsistency with earlier studies, but refine them by quantifying the
physically more relevant quantity of SF excess.

We have presented the average enhancement in SFR caused by major mergers
  of galaxies with masses above $10^{10} M_\sun$ at $0.4 <z<0.8$, deriving
  that approximately 8\% of the SF in the volume is {\it directly}
  triggered by major merging. As we have mentioned before, the SFR
  enhancement $\epsilon$ seems to be roughly independent
  of the quiescent SFR ground level present in the galaxy population, that is,
  insensitive to the drop in the SFR density of the Universe since
  z=1. If this is true, it means that the fraction of star formation directly
  triggered by galaxy interactions (given a mass cut) would depend only on the
  number of galaxies undergoing interactions. Using the evolution of pair
  fraction found in \citet{kart} we
   can infer a directly-triggered SF fraction of 1-2\% in the local Universe,
   as well as a fraction of 14-18\% at z=1. On the other hand, assuming no
   evolution in the pair fraction would keep the merger--triggered fraction at 8\% between $z=0$
   and $z=1$. These numbers have to be taken
   extremely carefully by the reader, as we present here only a crude
   extrapolation of our results to different redshifts in order to get an idea
   of the importance of the merger-driven star formation in the Universe.

There are a number of limitations of our result that should be borne in 
mind.
First, we can only include star formation rates $\ga 5\,M_\sun{\rm yr}^{-1}$ 
for $z\sim 0.6$ galaxies.  Therefore, our estimates of the SF contribution
from merging may be an upper limit because the merger-driven boost in SFR will cause
more objects to satisfy this criterion. Second, there are uncertainties in the conversion of 24{\micron}
to total IR, which could influence the excess star formation in 
close pairs or mergers by $\sim30\%$ \citep{papovich02,zheng07a}; this could be addressed once 
longer-wavelength, deep {\it Herschel} PACS observations become
available. We can only calculate an absolute upper limit by using the
  value of the enhancement found in the extreme case in which the 24\um to TIR
  conversion in all the
  interactions, independently of their luminosity, is calculated using an
  Arp220 template.  Using as input $\epsilon =3.1\pm 0.6$ for
  Eq.~\ref{eq:final} we would find a directly triggered fraction of $19\pm 5\%$,
  consistent with an scenario in
  which the underlying level of SF is basically negligible and most of the
  new stars are
  being formed in the burst mode. 
Third, it is conceivable that some enhanced star formation 
occurs in very late-stage merger remnants that were no longer recognized as
remnants and hence were not included 
in this census. This is both a practical (classification) and conceptual
issue: when does one declare a merger remnant a normal galaxy again?
Nonetheless, despite these points, the analysis presented here has made it very clear that 
only a small fraction of star formation in galaxies
with $M_* > 10^{10} M_\sun$ at $0.4<z<0.8$ is 
triggered by major interactions/mergers.

\subsection{Comparison with theoretical expectations}

Star formation enhancement in mergers has been
studied extensively with hydrodynamical N-body simulations
\citep[e.g., ][]{bh91,mh94,MH96,springel2000,cox06,cox08,dimatteo07}. However,
large-scale cosmological simulations lack the dynamic range to resolve the
internal dynamics of galaxies, crucial for modeling the gas inflows and the associated enhancement in
star formation. Therefore, the majority of these studies
\citep[except][]{tissera} have been of binary galaxy mergers with idealized
initial conditions, typically bulgeless or late-type disks. In most
studies, the properties of these progenitor disks are chosen to be
representative of present-day, relatively massive spiral galaxies such
as the Milky Way. These studies have shown that the burst efficiency
in mergers is sensitive to parameters such as merger mass ratio and
orbit, and progenitor gas fraction and bulge content. Therefore, any attempt
to use these results in an ensemble comparison 
must somehow convolve these dependencies with a redshift dependent,
cosmologically motivated distribution function for these quantities. In addition, \citet{cox06} have shown that star formation
enhancement in mergers can also depend on the treatment of supernova
feedback in the simulations. Furthermore, the detailed star formation
history during the course of a merger, particularly in the late
stages, may depend on the presence of an accreting supermassive black
hole \citep{dimatteo05}.

Let us consider the results from representative examples of such
binary merger simulations, by \citet{cox08}, who studied a broad range of merger mass ratios, gas fractions, and
progenitor B/T ratios, as well as exploring the
effects of two different SN feedback recipes. The 1:1 merger of two
``Milky Way''-like progenitors (shown in their Figure 12) shows an average factor of
$\sim 1.5$ enhancement in SF over about 2.5 Gyr, and a larger enhancement of a factor of 2--10
for a shorter period of about 0.6 Gyr. The overall average enhancement
over the whole merger is about a factor of 2.5, depending on the
precise timescale one averages over. Very large enhancements ($\sim 5-10$) occur over a very short timescale, $\lesssim
100$ Myr. This particular simulation represents the largest expected SF
enhancement, as the burst efficiency increases strongly towards equal merger
mass ratio. For mergers with 1:2.3 mass ratio (Figure 10 of \citealp{cox08}),
there is an enhancement of a factor of $\sim1.5$ for 2.5--3 Gyr, and of 2.5 for about 0.6 Gyr. It is also interesting to note that
the star formation rate in the late stages of the merger, when the
galaxy still appears morphologically disturbed (see Fig.~7 of \citealp{cox08})
is {\it depressed} with respect to the isolated case. A diverse set of progenitor morphologies, ranging from ellipticals to
late-type spirals, was studied by \citet{dimatteo07}. Overall, their results
are qualitatively similar to those from \citet{cox08}.

The simulations discussed so far aimed to reflect progenitor disks with gas
fractions, sizes, and morphologies typical of relatively massive,
low-redshift late-type spirals such as the Milky Way: gas fraction
$f_g \sim 0.2$; $B/T \sim 0.2$; scale length $r_d \sim 3$ kpc.
 However, \citet{hopkins08}
show that the burst efficiency is strong function of
progenitor gas fraction, in the sense that higher gas fraction
progenitors have {\it weaker} fractional enhancements. The burst efficiency is a factor
of eight lower for a gas fraction of 90 \% than for the
canonically-used value of 20\%. It is worth noting here that we find the
  same level of SF enhancement in major mergers at $z \sim 0.6$ and $z \sim
  0.1$, where the gas fractions of the two samples are expected to be rather
  different (see \S \ref{compar}). Whether or not this is quantitatively at odds with
  the expectations of \citet{hopkins08} remains to be seen. On the other
    hand, recent results from \citet{dimatteo08} show no difference between
    the strength or duration of tidally-triggered bursts of star formation in
    local Universe and their higher redshifts counterparts, in good agreement
    with the present study.

To place results in a cosmological context, \citet{somer08} used the results from a large
suite of hydrodynamic merger simulations \citep{cox06b,cox06c,rob1,rob2,rob3} to parameterize the dependence of burst efficiency and timescale on
merger mass ratio, gas fraction, progenitor circular velocity,
redshift, and the assumed effective equation of state. They
implemented these scalings within a cosmological semi-analytic merger
tree model. We applied our selection criteria to mock catalogs from
\citet{somer08}, by comparing the fraction of SFR produced in the triggered mode
in galaxies with $M_*>2\times 10^{10}M_{\sun}$ in our redshift range which suffered a major merger in the
last 500 Myr
with the total SFR occuring in galaxies $M_*>10^{10}M_{\sun}$. We found
that approximately 7\% of the SFR in the volume is produced in the burst mode
triggered by major mergers. This is in excellent agreement with the $8\pm3\%$
of the overall SFR being directly triggered by major interactions we showed in
the previous section.

\section{Conclusions} \label{conc}

To quantify the {\it average} 
effect
of major mergers on SFRs in galaxies, we 
have studied the enhancement of SF caused by major mergers 
between galaxies with $M_* > 10^{10} M_{\sun}$ at $0.4<z<0.8$. We combined
redshifts and stellar masses from COMBO-17 with high-resolution imaging based
on HST/ACS data for two fields (ECDFS/GEMS and A901/STAGES) and with star
formation rates that draw on UV and deep 24\um data from {\it Spitzer} to form
a sample a factor of two larger than previous studies in this redshift range. We then applied robust two-point correlation function techniques, supplemented by morphologically-classified very
close pairs and merger remnants to identify interacting galaxies. Our main findings are as follows:

\begin{enumerate}
\item Major mergers and interactions between star-forming massive galaxies trigger, on average, a mild 
enhancement in the SFR in pairs separated by
  projected distances $r_P \la 40$\,kpc; we find an enhancement of $\epsilon =
  1.80 \pm 0.30$ considering the SF-All cross-correlation, where only one
  galaxy in the pair is required to be forming stars. For a similar analysis
  using the autocorrelation of star forming galaxies we find $\epsilon = 1.50 \pm 0.25$.
\item Our results agree well with previous studies of SF enhancement 
  using close pairs at $z<1$. In particular, the behavior of 
SF enhancement at $z=0.1$, $z=0.6$ and $z=1$ appear to be rather similar, indicating that the average SFR
  enhancement in galaxy interactions is independent of the 
`pre-existing'
  SFR in the population.
\item We combine our estimate of the average SFR enhancements in major mergers
  with the global SFR to show that overall, $8 \pm 3\%$
  of the total star formation at these epochs is {\it directly} triggered by
  major interactions. We conclude that major mergers 
are an insignificant factor in stellar mass growth at $z<1$.
\item Major interactions do, however, play a key role in 
  triggering the most intense
  dust-obscured starbursts: We find that the majority of galaxies with
  IR-luminosities in excess of $3\times 10^{11}L_\sun$ are visually
  classified as ongoing mergers or found in projected pairs within 
$<40$\,kpc
  separation. This is not in disagreement with the small {\it average} SFR
  enhancement if the most intense SF bursts last only $\sim 100$\,Myr.
\item Our results for the SF enhancement appear to be in qualitative agreement with 
the extensive suite of hydrodynamical simulations by 
\cite{dimatteo07,dimatteo08} and \cite{cox08}, who produce both intense, short-lived bursts
of SF in some interactions, but yet produce average enhancements 
of only 25-50\% averaged over the $\sim 2$\,Gyr timescale
taken to complete the merger. Furthermore, we find excellent agreement between
the fraction of the total SFR directly triggered by major merging measured
here and the 7\% calculated from mock catalogues obtained from \citet{somer08}.

\end{enumerate}
\acknowledgements

We wish to thank an anonymous referee for the thorough report and many
comments which greatly improved this paper. We thank Cheng Li for sharing with us their results in the electronic
form. A.\ R.\ R.\ thanks Alejo Mart\'inez Sansigre and Arjen van der Wel for
productive discussions which helped to improve this paper. A.\ R.\ R.\ and E.\ F.\ B.\ gratefully acknowledge support through 
the Deutsche Forschungsgemeinschaft's Emmy Noether Programme. BH acknowledges
support by STFC. A.\ R.\ R.\ also
acknowledeges the Heidelberg--International Max Planck Research School program.

\end{document}